%% file: alsace_paper_arXiv.tex
\documentclass[a4paper,10pt]{article}

\usepackage{glossaries} 
\usepackage[utf8]{inputenc}
\usepackage[a4paper, total={7in, 8in}]{geometry}
\usepackage{siunitx}
\usepackage{algorithm, algorithmicx, algpseudocode}
\usepackage{amsmath,amsbsy}
\usepackage{booktabs}
\usepackage{amsfonts}
\usepackage{amsthm}
\usepackage{authblk}
\usepackage{bbm}
\usepackage{caption}
\usepackage{graphicx, subfigure}
\usepackage[colorlinks=true,bookmarksopen,bookmarksnumbered,linkcolor=black,citecolor=black,urlcolor=black]{hyperref}
\usepackage{mathtools}
\usepackage{multirow}
\usepackage{pgfplots}
\usepackage{tabularx}
\usepackage{tikz}
\usepackage{multirow}
\providecommand{\keywords}[1]{\textbf{\textit{keywords---}} #1}

\loadglsentries{abbreviations.tex}

\makeglossaries

\begin{document}

\title{Robust Adaptive Least Squares Polynomial Chaos Expansions in High-Frequency Applications}
\author[1,2,*]{Dimitrios Loukrezis}
\author[1]{Armin Galetzka}
\author[1,2]{Herbert De~Gersem}

\affil[1]{\small{Institut f\"ur Teilchenbeschleunigung und Elektromagnetische Felder (TEMF), Technische Universit\"at Darmstadt \protect\\ Schlossgartenstr. 8, 64289 Darmstadt, Germany}}
\affil[2]{\small{Centre for Computational Engineering, Technische Universit\"at Darmstadt \protect\\ Dolivostr. 15, 64293 Darmstadt, Germany}}
\affil[*]{{\small Corresponding author (loukrezis@temf.tu-darmstadt.de)}}
\date{}
\maketitle

\begin{abstract}
We present an algorithm for computing sparse, least squares-based polynomial chaos expansions, incorporating both adaptive polynomial bases and sequential experimental designs.
The algorithm is employed to approximate stochastic high-frequency electromagnetic models in a black-box way, in particular, given only a dataset of random parameter realizations and the corresponding observations regarding a quantity of interest, typically a scattering parameter.
The construction of the polynomial basis is based on a greedy, adaptive, sensitivity-related method.
The sequential expansion of the experimental design employs different optimality criteria, with respect to the algebraic form of the least squares problem.
We investigate how different conditions affect the robustness of the derived surrogate models, that is, how much the approximation accuracy varies given different experimental designs.
It is found that relatively optimistic criteria perform on average better than stricter ones, yielding superior approximation accuracies for equal dataset sizes.
However, the results of strict criteria are significantly more robust, as reduced variations regarding the approximation accuracy are obtained, over a range of experimental designs.
Two criteria are proposed for a good accuracy-robustness trade-off.

\end{abstract}

\section{Introduction}
\label{sec:intro}
For most, if not all, \gls{em} devices, \glspl{qoi} feature a parametric dependency upon the design characteristics of the device, e.g., its geometry or material properties.
During the design of an \gls{em} device, this dependency, denoted here with $g(\mathbf{y})$, where $\mathbf{y} \in \mathbb{R}^N$ is the parameter vector, is typically resolved with a computationally expensive parametric simulation, e.g., using a \gls{fe} model. 
In this work, our goal is to infer (learn, approximate) the relation between the design parameters of a high-frequency \gls{em} device and its \glspl{qoi}, e.g., one or more scattering parameters, and compute a black-box approximation $\widetilde{g} \approx g$, given only a dataset $\mathcal{D} = \left\{\mathbf{y}_l, g(\mathbf{y}_l)\right\}_{l=1}^L$.
This approximation is often called a surrogate model, a meta-model, or a response surface.
We call the set of parameter realizations $\left\{\mathbf{y}_l\right\}_{l=1}^L$ the \gls{ed} and the corresponding \gls{qoi} values $\left\{g(\mathbf{y}_l)\right\}_{l=1}^L$ the observations.
The latter shall here be simulation outputs, however, they could also refer to - possibly noisy - measurement data as well. 

The aforementioned inference problem is here considered in an \gls{uq} setting \cite{smith2014, sullivan2015}.
In particular, the model parameters are assumed to be independent \glspl{rv} $Y_n$, $n=1,\dots,N$, forming the $N$-variate \gls{rv} $\mathbf{Y} = \left(Y_1, \dots, Y_N\right)$.
The latter is defined on the probability space $\left(\Theta, \Sigma, P\right)$ and follows the \gls{pdf} $\varrho: \Xi \rightarrow \mathbb{R}_{\geq 0}$, where $\Xi$ denotes the image space.
Due to the \gls{rv} independence, it holds that $\varrho(\mathbf{y}) = \prod_{n=1}^N \varrho_n(y_n)$, where $\mathbf{y} = \mathbf{Y}\left(\theta\right) \in \Xi$, $\theta \in \Theta$, is now a \gls{rv} realization.
We note that \gls{rv} independence is not a crucial assumption and that dependencies can also be handled via suitable \gls{rv} transformations \cite{feinberg2018, jankoski2019, lebrun2009}.
The \glspl{rv} represent here random deviations from the specifications of a high-frequency device, which may arise due to manufacturing tolerances, material contamination, or other uncertainty sources.

Assuming that $g$ corresponds to a smooth functional relation, a computationally efficient approach for constructing a surrogate model is to compute a \gls{pce} \cite{ghanem1991, xiu2002}
\begin{equation}
\label{eq:gpc_1}
\widetilde{g}(\mathbf{y}) = \sum_{m=1}^M s_m \Psi_m(\mathbf{y}),
\end{equation}
where $s_m$ are scalar coefficients and $\Psi_m$ are polynomials orthogonal to the input \gls{pdf}.
Once available, the \gls{pce} can replace the original model in computationally demanding tasks, e.g., \gls{uq} or optimization studies.
For the purposes of \gls{uq}, certain statistical measures regarding the \gls{qoi} can be computed by simply post-processing the \gls{pce}'s terms \cite{blatman2010a, sudret2008}.
Moreover, the corresponding computational cost is typically orders of magnitude smaller than the one of a \gls{mc} method \cite{caflisch1998, lemieux2009}.
In this work, given a dataset $\mathcal{D}$ of size $L$, the \gls{pce} is constructed by solving the discrete \gls{ls} minimization problem
\begin{equation}
\label{eq:gpc_min_problem_1}
\widetilde{g} = \argmin_{\pi \in \mathbb{P}_M}  \sum_{l=1}^L \left(g\left(\mathbf{y}_l\right) - \pi\left(\mathbf{y}_l\right)\right)^2,
\end{equation}
where $\mathbb{P}_M = \text{span}\left\{\Psi_m, m=1,\dots,M\right\}$ denotes the corresponding polynomial space \cite{chkifa2015, migliorati2013, migliorati2014}.
Note that the surrogate model is constructed in a non-intrusive way, i.e., the model is used as a black box to compute the observations $g(\mathbf{y}_l)$, $l=1,\dots,L$.
The construction of the \gls{pce} can alternatively be based on compressive sensing \cite{diaz2018, doostan2011, hampton2015a, jakeman2015a, peng2014} or low-rank tensor decomposition methods \cite{doostan2013, konakli2016, konakli2016a}.
Nevertheless, many recent works on both the theoretical properties of \gls{ls} methods \cite{chkifa2015, cohen2013, cohen2018, migliorati2013, migliorati2014, migliorati2015, migliorati2015a} and on \gls{ls}-\gls{pce} algorithms \cite{blatman2010, blatman2010a, blatman2011, burnaev2017, fajraoui2017, hadigol2018, migliorati2013a} indicate that the interest in this approach remains active.
In the context of this work, a further reason for investigating and improving the \gls{ls}-\gls{pce} method is its popularity in the setting of \gls{em} simulations \cite{gladwin2019, hu2018, nguyen2016, offermann2015, prasad2016}.

The approximation accuracy of the \gls{pce} is crucially affected by the choice of the polynomial space $\mathbb{P}_M$.
This is especially relevant in high-dimensional approximations, due to the fact that the dimension of $\mathbb{P}_M$ grows very fast with the number of \glspl{rv}, which constitutes a manifestation of the so-called curse of dimensionality \cite{bellman1957}.
To mitigate this problem, a sparse albeit expressive polynomial basis must be constructed \cite{blatman2010, blatman2011, cohen2018, doostan2011, jakeman2015a, migliorati2013a, peng2014}.
The first contribution of this work is exactly in this direction. 
Specifically, we propose a greedy-adaptive algorithm for the construction of the \gls{pce} basis, which takes into consideration the sensitivity of the \gls{qoi} to the input \glspl{rv} and the corresponding \gls{pce} terms.

Another crucial aspect regarding the stability of the \gls{ls} problem \eqref{eq:gpc_min_problem_1} and the accuracy of its solution is the relation between the size of the polynomial basis and the size of the \gls{ed}, equivalently, of the dataset \cite{chkifa2015, cohen2013, migliorati2014}. 
At minimum, the \gls{ls} system must not be underdetermined, i.e., it must hold that $L \geq M$.
Consequently, considering an adaptively constructed \gls{pce} basis, the \gls{ed} must be sequentially expanded in order to meet the stability requirements.
Theoretical \gls{ls} stability criteria have been established in the literature \cite{chkifa2015, cohen2013, migliorati2014, migliorati2015} and have been used to form sequential \gls{ed} strategies  \cite{cohen2018, migliorati2013a}.
However, it has been observed that relaxed criteria typically result in more accurate approximations for equal costs \cite{chkifa2015, migliorati2013, migliorati2013a, migliorati2014, migliorati2015a}.
Therefore, most works resort to heuristic criteria regarding the dynamic relation between the polynomial basis and the dataset \cite{blatman2010, blatman2011}.
In the same vein, optimal \gls{ed} criteria have been considered recently \cite{burnaev2017, diaz2018, fajraoui2017, hadigol2018}.
In this case, optimality refers to selecting the best available realizations over a pool of candidate realizations to enhance the \gls{ed}.

Mostly, the aforementioned works focus on the accuracy of the \glspl{pce} derived with the proposed heuristic conditions regarding sequential \glspl{ed}. 
However, studies on the robustness of the approximation, i.e., to what extent different \glspl{ed} affect the accuracy of a surrogate model constructed with a specific method or adaptivity criterion, have not been sufficiently addressed in the literature so far. 
There lies the second contribution of this work, which aims to address the issue of robustness. 
Specifically, we examine different optimality conditions during the sequential expansion of the \gls{ed} and their impact on both the accuracy and the robustness of the resulting \glspl{pce}.
In combination with the proposed greedy-adaptive polynomial basis, a fully adaptive \gls{pce} algorithm is developed, where both the polynomial basis and the \gls{ed} are sequentially/adaptively expanded.

Our method is tested on two simulation models from the field of high-frequency electromagnetics. 
First, an academic test case is considered, employing a simple rectangular waveguide with dielectric filling \cite{loukrezis2019, loukrezisPhD} and featuring up to $15$ parameters.
Second, we apply the method to an optical grating coupler model \cite{pitelet2019} with up to $5$ parameters.
By including the frequency in the parameter vector, we are able to approximate not only the parametric dependence, but also the frequency response of the model, within a given frequency range.
Naturally, the frequency response must also correspond to a smooth functional, e.g., sharp resonances shall increase the computational cost of the method, or might even render it inapplicable altogether.
For both considered numerical examples, the suggested approach results in accurate surrogate models for comparably low dataset sizes.
We observe the influence of the different optimality criteria upon the accuracy and the robustness of the \glspl{pce}.
On the one hand, relaxed criteria result in - on average - more accurate surrogate models, which however vary significantly from one another for different \glspl{ed}.
On the other hand, stricter criteria yield approximations of inferior accuracy, however, the variance of the approximation accuracy for different \glspl{ed} is significantly reduced.
Two sequential \gls{ed} criteria are identified, for which the trade-off between accuracy and robustness can be considered as acceptable.

The rest of this paper is organized as follows. In Section~\ref{sec:pce} we introduce the \gls{pce} as well as the computation of the corresponding coefficients via discrete \gls{ls}. This is followed by Section~\ref{sec:apce} where we present a scheme which exploits the sensitivity of the \gls{qoi} on the \glspl{rv} for adaptively selecting the \gls{pce} basis terms. In the same section, we extend the scheme by robust sequential \gls{ed}, relying on different optimality criteria. Numerical experiments on two high-frequency electromagnetic devices verify the reliability and accuracy of the presented method in Section~\ref{sec:applications}. Concluding remarks and possible continuations of this work are available in Section~\ref{sec:conclusion}.

\section{Least Squares Polynomial Chaos Expansions}
\label{sec:pce}
\subsection{Univariate Polynomial Chaos Expansions}
\label{subsec:uv_pce}
We first consider a univariate model $g(y)$, where $y = Y\left(\theta\right)$, $\theta \in \Theta$, and the \gls{rv} $Y$ is characterized by the \gls{pdf} $\varrho(y)$. 
We denote a univariate polynomial of degree $p \in \mathbb{Z}_{\geq 0}$ with $\psi_p$ and demand that the polynomial basis $\left\{\psi_p\right\}_{p=0}^{p_{\text{max}}}$ is orthogonal with respect to the univariate \gls{pdf}, such that
\begin{equation}
\label{eq:orth1d}
\mathbb{E}\left[\psi_p \psi_q\right] = \int_{\ImageSet} \psi_p(y) \psi_q(y) \varrho(y) \, \mathrm{d}y = \gamma_p \delta_{p,q}, 
\end{equation}
where $p,q \in \left\{0, 1, \dots, p_{\text{max}}\right\}$, $\delta_{p,q}$ is the Kronecker delta, and $\gamma_p$ a normalization factor.
In the rest of this paper, we will always assume that $\gamma_p=1$, $\forall p \in \left\{0, 1, \dots, p_{\text{max}}\right\}$, i.e., that $\left\{\psi_p\right\}_{p=0}^{p_{\text{max}}}$ is an orthonormal basis. 
For commonly used \glspl{pdf}, the \gls{gpc} or Wiener-Askey scheme \cite{xiu2002} provides correspondences to families of orthogonal polynomials.
Extensions to arbitrary \glspl{pdf} have been introduced by using numerically constructed orthogonal polynomials \cite{oladyshkin2012, soize2004, wan2006a}.
The univariate \gls{pce} reads
\begin{equation}
\label{eq:gpc1d}
g(y) \approx \widetilde{g}(y) = \sum_{p=0}^{p_{\max}} s_p \psi_p(y),
\end{equation}
where $s_p \in \mathbb{R}$ are scalar coefficients.
In essence, $\widetilde{g}$ is a polynomial living in the space
\begin{equation}
\mathbb{P}_{p_{\text{max}}} = \text{span}\left\{ \psi_p : p \leq p_{\text{max}} \right\}.
\end{equation}

\subsection{Multivariate Polynomial Chaos Expansions}
\label{subsec:mv_pce}
We proceed to the case of a multivariate model $g(\mathbf{y})$, where $\mathbf{y} = \mathbf{Y}\left(\theta\right)$ and the \glspl{rv} $\mathbf{Y}$ are characterized by the \gls{pdf} $\varrho(\mathbf{y}) = \varrho_1(y_1) \cdots \varrho_N(y_N)$. 
We introduce the multi-index $\mathbf{p} = \left(p_1, \dots, p_N\right) \in \mathbb{Z}_{\geq 0}^N$ which contains the polynomial order per parameter and defines the corresponding multivariate polynomial $\Psi_{\mathbf{p}}$ as
\begin{equation}
\label{eq:polyNd}
\Psi_\mathbf{p}(\mathbf{y}) = \prod_{n=1}^N \psi_{p_n} (y_n).
\end{equation}
In this case, the orthonormality condition reads
\begin{equation}
\label{eq:orthNd}
\mathbb{E}\left[\Psi_{\mathbf{p}} \Psi_{\mathbf{q}}\right] = \int_\ImageSet \Psi_{\mathbf{p}}\left(\mathbf{y}\right) \Psi_{\mathbf{q}}\left(\mathbf{y}\right) \varrho\left(\mathbf{y}\right) \mathrm{d}\mathbf{y} =  \delta_{\mathbf{p}\mathbf{q}},
\end{equation}
where $\delta_{\mathbf{p}\mathbf{q}} = \delta_{p_1q_1} \cdots \delta_{p_Nq_N}$.
Assuming a polynomial basis $\left\{\Psi_{\mathbf{p}}\right\}_{\mathbf{p} \in \Lambda}$, where  $\Lambda$ is a multi-index set, the multivariate \gls{pce} reads
\begin{equation}
\label{eq:gpcNd}
g\left(\mathbf{y}\right) \approx \widetilde{g}\left(\mathbf{y}\right) =  
\sum_{\mathbf{p} : \mathbf{p} \in \Lambda} s_{\mathbf{p}} \Psi_{\mathbf{p}}\left(\mathbf{y}\right),
\end{equation} 
and the corresponding multivariate polynomial space $\mathbb{P}_\Lambda$ is given by
\begin{equation}
\label{eq:poly_space_gpc_Nd}
\mathbb{P}_\Lambda = \text{span}\left\{\Psi_{\mathbf{p}} : \mathbf{p} \in \Lambda \right\}.
\end{equation}
Common choices for $\Lambda$ in the literature are \gls{tp}, \gls{td}, \gls{hc}, and \gls{dc} multi-index sets \cite{babuska2010, chkifa2015, migliorati2014}. 
The respective definitions are given in Table~\ref{tab:lambdas}, where $\mathbf{e}_n$ denotes the $n$-th unit vector.
We note that multi-index sets of arbitrary shapes may also be used, see, e.g., \cite{blatman2010, blatman2010a}. 

\begin{table}[t!]
	\caption{Definitions of commonly employed multi-index sets, respectively, polynomial bases.}
	\centering
	\setlength{\tabcolsep}{1em} 
	{\renewcommand{\arraystretch}{1.5}
		\begin{tabular}{@{}ll@{}}
			\toprule
			TP & $\Lambda^\mathrm{TP} \coloneqq \left\{\mathbf{p} : \max_n(p_n) \leq p^{\max}, p^{\max} \in \mathbb{Z}_{\geq 0} \right\}$ \\
			TD & $\Lambda^\mathrm{TD} \coloneqq \left\{\mathbf{p} : \sum_{n=1}^N p_n \leq p^{\max}, p^{\max} \in \mathbb{Z}_{\geq 0} \right\}$ \\
			HC & $\Lambda^\mathrm{HC} \coloneqq \left\{\mathbf{p} : \prod_{n=1}^N \left(p_n+1\right) \leq p^{\max}+1, p^{\max} \in \mathbb{Z}_{\geq 0} \right\}$ \\
			DC & $\Lambda^\mathrm{DC} \coloneqq \left\{\mathbf{p} : \left(\mathbf{p} - \mathbf{e}_n\right) \in \Lambda^\mathrm{DC}, \forall n=1,\dots,N \: \text{with} \: p_n>0 \right\}$ \\ 
			\bottomrule
		\end{tabular}
	}
	\label{tab:lambdas}
\end{table}

\subsection{Computing Expansion Coefficients via Discrete Least Squares}
\label{subsec:gpc_coeffs}
We now assume that a polynomial basis $\left\{\Psi_{\mathbf{p}}\right\}_{\mathbf{p} \in \Lambda}$ with cardinality $\#\Lambda = M$, as well as an \gls{ed} $\left\{\mathbf{y}_l\right\}_{l=1}^L$ and the corresponding observations $\left\{g\left(\mathbf{y}_l\right)\right\}_{l=1}^L$, are available.
Then, the \gls{pce} can be obtained by solving the minimization problem \eqref{eq:gpc_min_problem_1}, where the polynomial space is defined as in \eqref{eq:poly_space_gpc_Nd}.
For the solution of \eqref{eq:gpc_min_problem_1}, we introduce the design matrix $\mathbf{D} \in \mathbb{R}^{L \times M}$ with elements $d_{lm}=\Psi_{m}\left(\mathbf{y}_l\right)$, and the observation vector  $\mathbf{b}=\left(g(\mathbf{y}_1), \dots, g(\mathbf{y}_L)\right)^\top$. 
Collecting the unknown \gls{pce} coefficients into a vector $\mathbf{s} = \left(s_1, \dots, s_M\right)^\top$, we form the discrete minimization problem
\begin{equation}
\label{eq:min_algebraic}
\mathbf{s}=\argmin_{\hat{\mathbf{s}} \in \mathbb{R}^M} \|\mathbf{D} \hat{\mathbf{s}} - \mathbf{b} \|_2.
\end{equation}
Applying the necessary conditions for a minimum, we obtain the normal equation \cite[Section 20.4]{higham2002}
\begin{equation}
\label{eq:normal_solution}
\left(\mathbf{D}^\top \mathbf{D}\right) \mathbf{s}  =  \mathbf{D}^\top \mathbf{b},
\end{equation}
where the system matrix $\mathbf{A} = \mathbf{D}^\top \mathbf{D}$ is called the information matrix. 
The solution to \eqref{eq:normal_solution} is unique if the design matrix is nonsingular.
Moreover, it must obviously hold that $L \geq M$, i.e., the system of equations cannot be underdetermined.

Due to the well-known estimates regarding the sensitivity of the \gls{ls} solution and its dependence on the condition number $\kappa\left(\cdot\right)$ of the corresponding system matrix \cite{golub1996, higham2002}, it is generally not recommended to use the normal equation \eqref{eq:normal_solution}, as it can easily be shown that 
\begin{equation}
\label{eq:cond_problem}
\kappa\left(\mathbf{A}\right) = \left(\kappa\left(\mathbf{D}\right)\right)^2.
\end{equation}
A QR decomposition of the design matrix $\mathbf{D}$ can be used instead \cite[Section 20.2]{higham2002}, in which case the conditioning of the \gls{ls} system is given by $\kappa\left(\mathbf{D}\right) = \kappa\left(\mathbf{Q} \mathbf{R}\right) = \kappa\left(\mathbf{R}\right)$.

\section{Adaptive Least Squares Polynomial Approximations}
\label{sec:apce}
\subsection{Adaptive Polynomial Basis}
\label{subsec:adapt_bases}
We first address the case of a fixed dataset $\mathcal{D} = \left\{\mathbf{y}_l, b_l\right\}_{l=1}^L$, which is considered to be sufficient for computing a \gls{pce} with $M$ terms.
The question which arises is, which $M$ polynomials, equivalently, which multi-index set $\Lambda$ with $\#\Lambda = M$ will result in an accurate approximation.
Several algorithms have been developed to address this problem \cite{blatman2010, blatman2010a, blatman2011, doostan2011, jakeman2015a, hadigol2018, hampton2015, migliorati2013a, peng2014}.
Any of these approaches can be combined with the sequential \gls{ed} strategies discussed in Section~\ref{subsec:adapt_ed}.

Additionally to the aforementioned methods, we present here yet another algorithm for the adaptive construction of the \gls{pce} basis.
Our approach is conceptually similar to a well-known dimension-adaptive quadrature method \cite{gerstner2003}, therefore, we enforce the use of \gls{dc} multi-index sets, as defined in Table~\ref{tab:lambdas}.
We note that \gls{dc} sets are not strictly necessary.
For example, the simple two-dimensional function
\begin{equation}
\label{eq:pathological}
g\left(y_1, y_2\right) = a y_1^2 + b y_2,
\end{equation}
can be represented exactly by a \gls{pce} based on  $\Lambda = \left\{ \left(2,0\right), \left(0,1\right) \right\}$. 
However, the \gls{dc} property would require that $\Lambda = \left\{ \left(0,0\right), \left(1,0\right), \left(2,0\right), \left(0,1\right) \right\}$, thus unnecessarily augmenting the \gls{ls} system matrix.
Nevertheless, while not optimal, \gls{dc} sets are employed in several theoretical works \cite{chkifa2015, cohen2017, cohen2018, migliorati2013a, migliorati2013, migliorati2014} due to the fact that the corresponding polynomial spaces satisfy a number of desirable properties, e.g., closure under differentiation for any variable, and invariance under a change of basis.
Moreover, as also verified by the results in Section~\ref{sec:applications}, \glspl{pce} based on \gls{dc} sets perform very well in practice. 
This can be attributed to the fact that pathological cases such as \eqref{eq:pathological} are rarely encountered in practical applications.

The adaptive construction of the \gls{pce} basis proceeds as follows.
Let us assume that a multivariate approximation \eqref{eq:gpcNd} based on a \gls{dc} multi-index set $\Lambda$ is readily available.
If not, we can initialize the procedure with $\Lambda = \left\{\left(0,0,\dots,0\right)\right\}$.
We call ``admissible neighbors'' the indices which do not belong to $\Lambda$ and would satisfy the \gls{dc} property if added to $\Lambda$.
The corresponding admissible set is defined as
\begin{equation}
\label{eq:admset}
\AdmSet \coloneqq \left\{ \mathbf{p} \not \in \Lambda \, : \, \left(\mathbf{p}-\mathbf{e}_n\right) \subset  \Lambda, \forall n=1,\dots,N \:\: \text{with} \:\: p_n>0 \right\}.
\end{equation}
Next, we construct the basis corresponding to the multi-index set $\Lambda^{\text{LS}} = \Lambda \cup \AdmSet$ and solve the discrete \gls{ls} minimization problem \eqref{eq:min_algebraic} for the coefficients $s_{\mathbf{p}}$, $\mathbf{p} \in \Lambda^{\text{LS}}$.
Assuming that orthonormal polynomials are used, the value $s_{\mathbf{p}}^2$ is the equivalent of the partial variance due to the multi-index $\mathbf{p}$, thus, directly linked to the contribution of that multi-index to the total variance of the \gls{qoi} \cite{blatman2010a, sudret2008}.
Since Sobol sensitivity indices are nothing more than fractions of partial variances over the total variance of the \gls{qoi}, the value $s_{\mathbf{p}}^2$ can be interpreted as a sensitivity indicator regarding the multi-index $\mathbf{p}$. 
Therefore, we add to $\Lambda$ the admissible multi-index which corresponds to the maximum sensitivity indicator, such that
\begin{equation}
\Lambda \leftarrow \Lambda \cup \left\{\mathbf{p}^{*}\right\}, \hspace{0.5em} \text{where} \hspace{0.5em} \mathbf{p}^{*} = \argmax_{\mathbf{p} \in \AdmSet} s_{\mathbf{p}}^2.
\end{equation}
This procedure continues iteratively until $\#\Lambda = M$ basis terms are reached, as shown in Algorithm~\ref{algo:adapt_basis}.

\begin{algorithm}[h!]
	\begin{algorithmic}
		\State \textbf{Data}: dataset $\mathcal{D}$, maximum PCE terms $M$, initial \gls{dc} multi-index set $\Lambda^{\text{init}}.$ 
		\State \textbf{Result}: \gls{dc} multi-index set $\Lambda$ with $\#\Lambda=M$, PCE basis $\left\{\Psi_{\mathbf{p}}\right\}_{\mathbf{p} \in \Lambda}$ and coefficients $\left\{s_{\mathbf{p}}\right\}_{\mathbf{p} \in \Lambda}$.
		\While {$\#\Lambda < M$}
		\State Create the extended multi-index set $\Lambda^{\text{LS}} = \Lambda \cup \Lambda^{\text{adm}}$. 
		\State Solve the LS problem \eqref{eq:min_algebraic} using $\Lambda^{\text{LS}}$.  
		\State Find the admissible multi-index corresponding to the largest sensitivity indicator, i.e., $\mathbf{p}^{*} = \argmax_{\mathbf{p} \in \AdmSet} s_{\mathbf{p}}^2$. 
		\State Expand $\Lambda$ with $\mathbf{p}^*$, i.e., $\Lambda = \Lambda \cup \mathbf{p}^*$. 
		\EndWhile
	\end{algorithmic}
	\caption{Adaptive PCE basis construction.} 
	\label{algo:adapt_basis}
\end{algorithm}

\subsection{Sequential Experimental Design}
\label{subsec:adapt_ed}
We now consider the case where the \gls{pce} is expanded adaptively until it reaches a desired accuracy $e$.
This accuracy is typically estimated using a \gls{cv} error metric, e.g., the \gls{loo} \cite{blatman2010a, blatman2011, fajraoui2017} or the $\ell^{\infty}$ \cite{migliorati2013, migliorati2013a, migliorati2014} \gls{cv} error.
In that case, the dataset $\mathcal{D}$, accordingly, the \gls{ed} and the observations, should also be expanded, such that the \gls{ls} problem remains stable and the accuracy of the \gls{pce} increases.
Moreover, this expansion should be sequential, such that previously available observations can be re-used, thus restricting the computational cost to the simulations due to the new parameter realizations.
Such an approach falls in the category of sequential \glspl{ed} \cite{arras2019, blatman2010, blatman2010a, blatman2011, fajraoui2017}. 

We will focus here on three such criteria. 
Denoting with $\mathbf{G} = L^{-1} \mathbf{A}$ the normalized information matrix, those criteria are, (i) the $K$-optimality criterion, which aims to minimize the condition number $\kappa\left(\mathbf{G}\right)$, (ii) the $A$-optimality criterion, which aims to minimize the trace $\mathrm{tr}\left(\mathbf{G}^{-1}\right)$, and (iii) the $E$-optimality criterion, which aims to minimize the maximum eigenvalue $\lambda_{\max}\left(\mathbf{G}^{-1}\right)$.
We note that $\kappa\left(\mathbf{G}\right) = \kappa\left(\mathbf{A}\right) = \kappa\left(\mathbf{D}\right)^2$,  therefore, the $K$-optimality criterion can be modified such that  the design matrix $\mathbf{D}$ is used instead.
Moreover, it can be easily shown that $\lambda_{\max}\left(\mathbf{G}^{-1}\right) = \left(\lambda_{\min}\left(\mathbf{G}\right)\right)^{-1}$, thus, we can avoid the possibly costly inversions.
The inversion can also be avoided when the trace-based criterion is used, by exploiting the property $\mathrm{tr}\left(\mathbf{G}^{-1}\right) = \sum_{m=1}^M \left(\lambda_{m}\left(G\right)\right)^{-1}$, where $\lambda_{m}\left(G\right)$ denotes the $m$-th eigenvalue of matrix $\mathbf{G}$.

Contrary to the setting of optimal \glspl{ed}, in this work we do not seek to minimize those measures. 
Instead, we investigate conditions which, if violated, trigger the expansion of the dataset, equivalently, of the \gls{ed} and the observations.
In essence, we enforce the values of $\kappa\left(\mathbf{D}\right)$, $\mathrm{tr}\left(\mathbf{G}^{-1}\right)$, $\lambda_{\max}\left(\mathbf{G}^{-1}\right)$ to be below some limit value. 
If this condition is satisfied, the \gls{pce} is adaptively expanded using the available \gls{ed}, as in Section~\ref{subsec:adapt_bases}.
Otherwise, the polynomial basis remains fixed and the dataset is expanded until the condition is again satisfied.
This sequential \gls{ed} strategy is depicted in Algorithm~\ref{algo:seq_ed}.
Several works claim that relaxed conditions may lead to more accurate \glspl{pce} for equal costs, equivalently, for \glspl{ed} of equal sizes \cite{chkifa2015, migliorati2013, migliorati2013a, migliorati2014}.
However, as we will show in Section~\ref{sec:applications}, this accuracy improvement comes at the cost of robustness, in the sense that the accuracy of the \gls{pce} depends significantly on the available \gls{ed}.
This aspect has not received much attention in the literature so far. 

\begin{algorithm}[h!]
	\begin{algorithmic}
		\State \textbf{Data}: initial dataset $\mathcal{D}^\text{init}$, initial multi-index set $\Lambda^{\text{init}}$, desired accuracy $e$.
		\State \textbf{Result}: final multi-index set $\Lambda$, PCE basis $\left\{\Psi_{\mathbf{p}}\right\}_{\mathbf{p} \in \Lambda}$, and coefficients $\left\{s_{\mathbf{p}}\right\}_{\mathbf{p} \in \Lambda}$.
		\State $\Lambda = \Lambda^{\text{init}}$, $\mathcal{D} = \mathcal{D}^{\text{init}}$. 
		\While {desired accuracy is not reached}
		\While {$\kappa\left(\mathbf{D}\right) \leq \kappa^{\mathrm{limit}}$ or $\mathrm{tr}\left(\mathbf{G}^{-1}\right) \leq \mathrm{tr}^{\mathrm{limit}}$ or $\lambda_{\max}\left(\mathbf{G}^{-1}\right) \leq \lambda_{\max}^{\mathrm{limit}}$}
		\State Expand the PCE using Algorithm~\ref{algo:adapt_basis}.
		\EndWhile
		\State Expand the dataset $\mathcal{D}$, equivalently, expand the ED and the QoI evaluations.
		\EndWhile
		
	\end{algorithmic}
	\caption{Sequential ED strategy.} 
	\label{algo:seq_ed}
\end{algorithm}

\section{Application Examples}
\label{sec:applications}
\subsection{Verification Methodology}
\label{subsec:verification}
In the following, we employ Algorithms~\ref{algo:adapt_basis} and \ref{algo:seq_ed} to approximate the input-output relation of two stochastic high-frequency models via \glspl{pce}.
The two models feature up to $15$ and $5$ uniform input \glspl{rv}, respectively.
Both algorithms are part of the in-house developed software ALSACE (Approximations via Least Squares Adaptive Chaos Expansions)\footnote{https://github.com/dlouk/ALSACE}, which is partially based on the OpenTURNS C++/Python library \cite{baudin2017}.

We compute \glspl{pce} using different criteria for the sequential expansion of the \gls{ed} shown in Algorithm~\ref{algo:seq_ed}.
For each criterion, we use multiple \glspl{ed} for the construction of the \gls{pce} and measure both the average approximation accuracy, as well as the variations around that mean accuracy value.
For a \gls{pce} computed with a specific \gls{ed}, the approximation accuracy is measured using the \gls{rms} \gls{cv} error
\begin{equation}
\label{eq:cverr_rms}
e_{\text{cv}, \text{RMS}} = \sqrt{\frac{1}{Q} \sum_{q=1}^Q \left(\widetilde{g}\left(\mathbf{y}_q\right) - g\left(\mathbf{y}_q\right)\right)^2},
\end{equation}
where a \gls{cv} sample $\left\{\mathbf{y}_q\right\}_{q=1}^Q$, which is randomly drawn from the joint input \gls{pdf}, is used.
We note that the \gls{cv} sample does not coincide with the \gls{ed}.

\subsection{Rectangular Waveguide with Dielectric Inset}
\label{subsec:diel_slab_wg}
\begin{figure}[t]
	\minipage[b][][b]{0.48\textwidth}
	\centering
	\def\svgwidth{0.6\textwidth}
	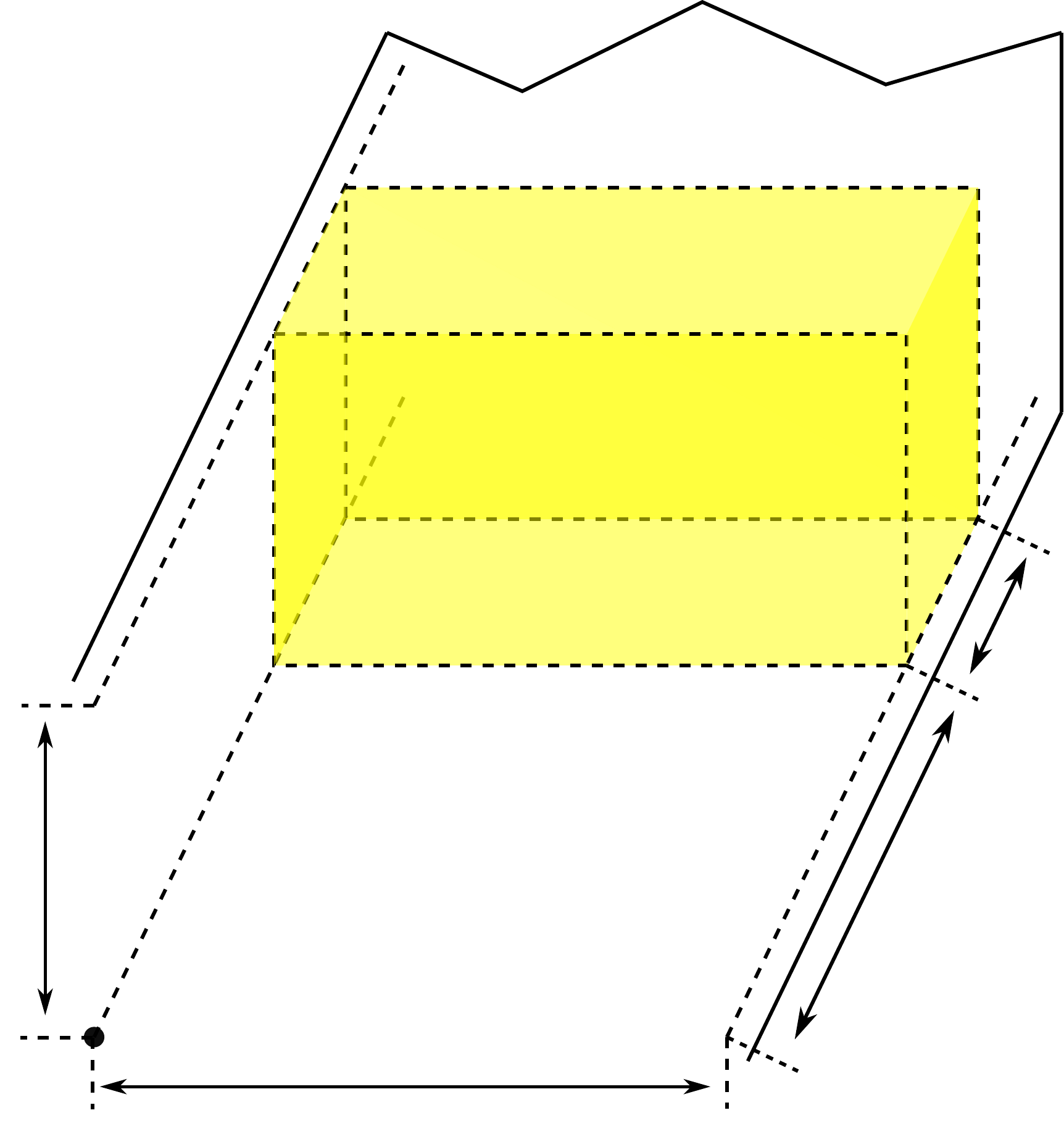
	\caption{Rectangular waveguide with dielectric inset.}
	\label{fig:diel_slab_wg}
	\endminipage\hfill
	\minipage[b][][b]{0.48\textwidth}
	\includegraphics{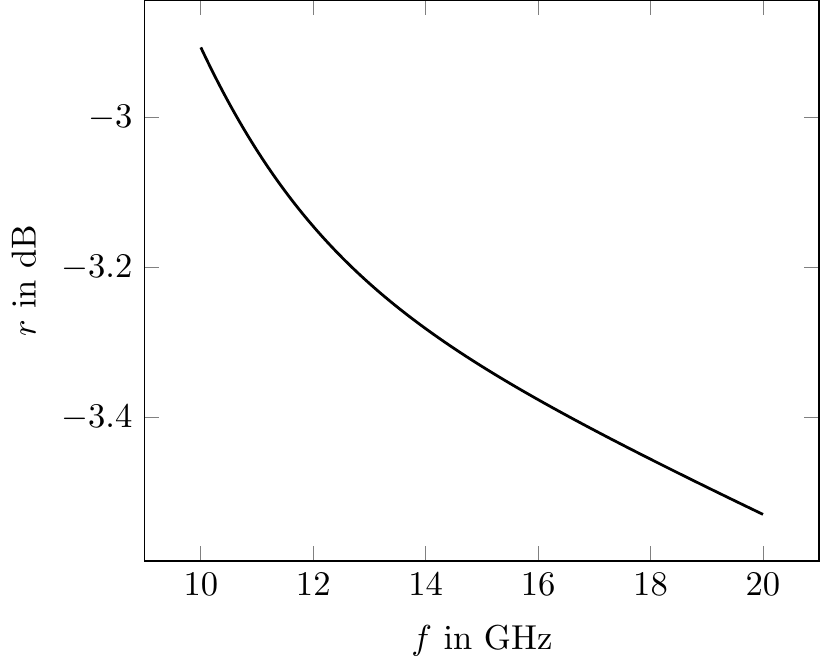} 
	\caption{Frequency-response of the dielectric-inset waveguide for the nominal geometry and material parameters.}
	\label{fig:debye2_broadband}
	\endminipage\hfill		
\end{figure}

As a first test case, we consider a rectangular waveguide with dielectric filling, as shown in Figure~\ref{fig:diel_slab_wg}.
The waveguide has width $w$, height $h$, and is infinitely extended in the positive $z$-direction. 
An incoming plane wave excites the waveguide at the input port boundary $\Gamma_\text{\text{in}}$.
For simplicity, the excitation coincides with the fundamental \gls{te} mode only, while all other propagation modes attenuate quickly in the structure.
The output port $\Gamma_{\text{out}}$, which is not shown in Figure~\ref{fig:diel_slab_wg}, is placed at a distance $d + \ell + d$ from $\Gamma_{\text{in}}$, where $\ell$ is the length of the dielectric material, placed at a distance $d$ from $\Gamma_\text{\text{in}}$.
The remaining waveguide walls are assumed to be \glspl{pec} and the corresponding boundary is denoted with $\Gamma_{\text{PEC}}$. 

The dielectric material has a permittivity $\epsilon = \epsilon_0 \epsilon_\text{r}$ and permeability $\mu = \mu_0 \mu_\text{r}$, where ``0'' denotes the property value in the free space and ``r'' its relative value in the material.
The relative material values are given by Debye relaxation models of second order \cite{xu2010}, such that
\begin{align}
\epsilon_{\mathrm{r}} &= \epsilon_{\infty} + \frac{\epsilon_{\mathrm{s},1} - \varepsilon_{\infty}}{1 + \left(\imath \omega \tau_{\epsilon,1}\right)} + \frac{\epsilon_{\mathrm{s},2} - \varepsilon_{\infty}}{1 + \left(\imath \omega \tau_{\epsilon,2}\right)} \\  
\mu_{\mathrm{r}} &= \mu_{\infty} + \frac{\mu_{\mathrm{s},1} - \mu_{\infty}}{1 + \left(\imath \omega \tau_{\mu,1}\right)} + \frac{\mu_{\mathrm{s},2} - \mu_{\infty}}{1 + \left(\imath \omega \tau_{\mu,2}\right)},
\end{align}
where $\tau_{\epsilon/\mu,1/2}$ are relaxation time constants, the subscript ``$\infty$'' refers to a very high frequency value of the relative material property, the subscript ``s'' to a static value of the relative material property, and $\imath$ denotes the imaginary unit.

Let $D$ be the computational domain, $\omega = 2 \pi f$ the angular frequency, $f$ the frequency, $\mathbf{E}$ the electric field, $\mathbf{U}^{\text{i}}$ the incoming plane wave, $\mathbf{n}$ the outwards-pointing normal vectors and $\mathbf{k} = \left(0,0,k_z\right)$ the wavevector.
Then, the underlying mathematical model reads
\begin{subequations}
	\label{eq:maxwell_source}
	\begin{align}
	\nabla \times \left( \mu^{-1}\nabla \times \mathbf{E}\right) - \omega^2 \epsilon \mathbf{E} &= 0, \quad &&\text{~in~} D,\\
	\mathbf{n}_{\Gamma_{\text{PEC}}} \times \mathbf{E} &= 0, \quad  &&\text{~on~} \Gamma_{\text{PEC}}, \\
	\mathbf{n}_{\Gamma_{\text{in}}} \times \left(\nabla \times \mathbf{E}\right) + \imath k_z \mathbf{n}_{\Gamma_{\text{in}}} \times \left(\mathbf{n}_{\Gamma_{\text{in}}} \times \mathbf{E}\right) &= \mathbf{U}^{\text{i}}, \quad  &&\text{~on~} \Gamma_{\text{in}}, \\
	\mathbf{n}_{\Gamma_{\text{out}}} \times \left(\nabla \times \mathbf{E}\right) + \imath k_z \mathbf{n}_{\Gamma_{\text{out}}} \times \left(\mathbf{n}_{\Gamma_{\text{out}}} \times \mathbf{E}\right) &= 0, \quad  &&\text{~on~} \Gamma_{\text{out}}.
	\end{align}
	\label{eq:Maxwell_boundary_problem}
\end{subequations}
The \gls{qoi} is chosen to be the reflection coefficient at the input port $\Gamma_\text{\text{in}}$,  $r = \left| \frac{\mathbf{E}_{\Gamma_{\text{in}}^-}}{\mathbf{E}_{\Gamma_{\text{in}}^+}}\right| \in \left[0,1\right]$. 
Usually, problem \eqref{eq:maxwell_source} is solved numerically, e.g., using the \gls{fem}. 
For this simple model, an analytical solution exists for the reflection coefficient $r$. 
Therefore, errors due to spatial discretization can be neglected and we can focus on the error due to the truncation of the \gls{pce} alone.

We introduce uncertainties with respect to all geometrical and Debye material model parameters, and collect them in a $14$-dimensional random vector $\mathbf{Y}$.
In the nominal configuration, the parameter values are $\bar{w}=\SI{30}{\milli\meter}$, $\bar{h}=\SI{3}{\milli\meter}$, $\bar{\ell}=\SI{7}{\milli\meter}$, $\bar{d}=\SI{1}{\milli\meter}$, $\bar{\epsilon}_{\text{s},1}=2$, $\bar{\epsilon}_{\text{s},2}=2.2$, $\bar{\epsilon}_{\infty}=1$,  $\bar{\mu}_{\text{s},1}=2$, $\bar{\mu}_{\text{s},2}=3$, $\bar{\mu}_{\infty}=1$, $\bar{\tau}_{\epsilon,1} = 1$, $\bar{\tau}_{\epsilon,2} = 1.1$, $\bar{\tau}_{\mu,1} = 1$, and $\bar{\tau}_{\mu,1} = 2$.
Each parameter is now assumed to follow a uniform distribution with bounds given by $\bar{y}_n \pm 0.05 \bar{y}_n$, i.e., a uniform random variation around the nominal value up to a maximum of $5\%$ is introduced.
Denoting with $\mathbf{y} = \mathbf{Y}\left(\theta\right)$ a realization of the random vector $\mathbf{Y}$, the parametric counterpart of problem  \eqref{eq:Maxwell_boundary_problem} features parameter-dependent material properties $\epsilon\left(\mathbf{y}\right)$, $\mu\left(\mathbf{y}\right)$, computational domain $D\left(\mathbf{y}\right)$, and boundaries $\Gamma_{\text{PEC}}\left(\mathbf{y}\right)$, $\Gamma_{\text{in}}\left(\mathbf{y}\right)$, $\Gamma_{\text{out}}\left(\mathbf{y}\right)$.
Accordingly, the field solution $\mathbf{E}\left(\mathbf{y}\right)$ and the reflection coefficient $r\left(\mathbf{y}\right)$ are parameter-dependent as well.

\subsubsection{Single-Frequency Surrogate Modeling}
\label{subsubsec:debye2_sf}
\begin{figure}[t]
	\begin{center}
		\includegraphics[width=0.339\textwidth]{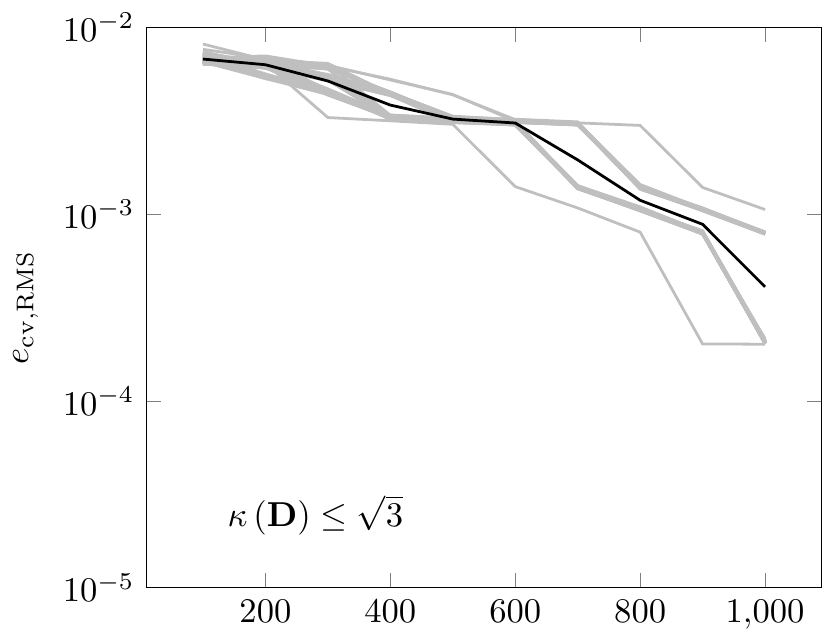} 
		\hfill
		\includegraphics[width=0.32\textwidth]{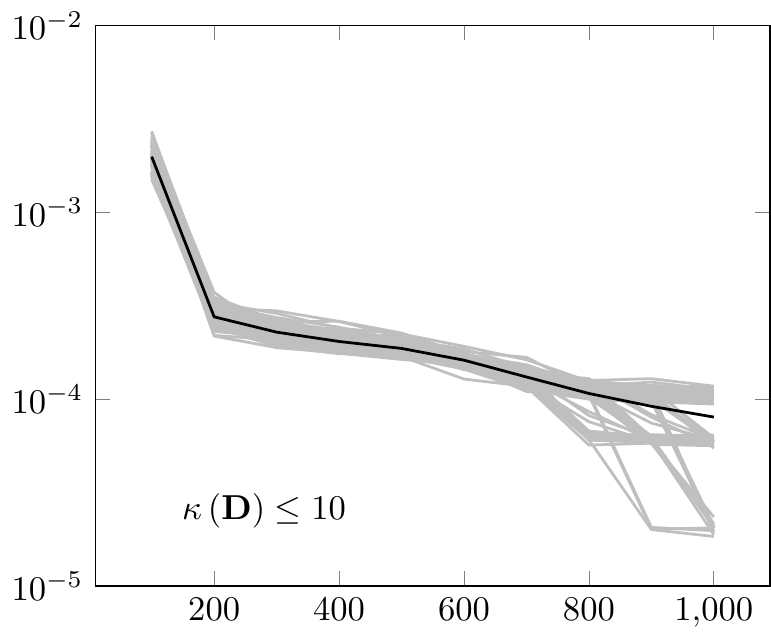}
		\hfill
		\includegraphics[width=0.32\textwidth]{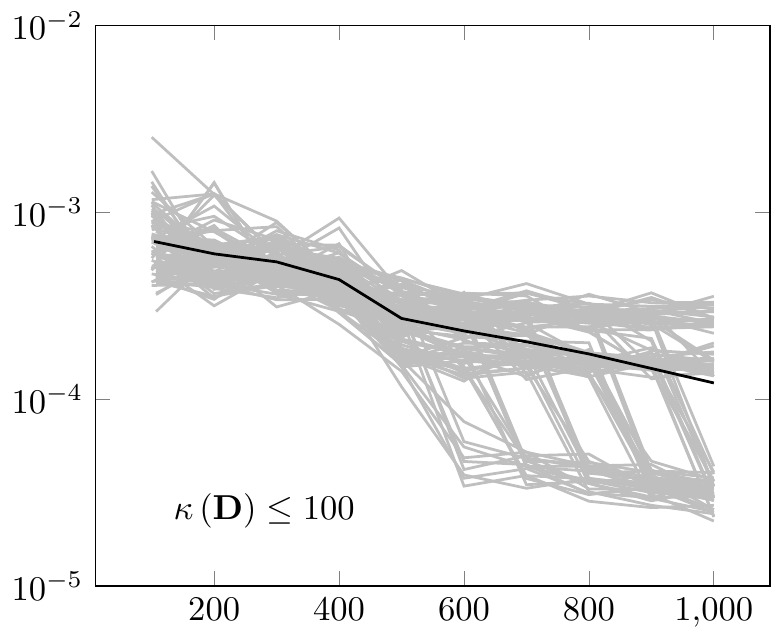} 
		\\
		\includegraphics[width=0.338\textwidth]{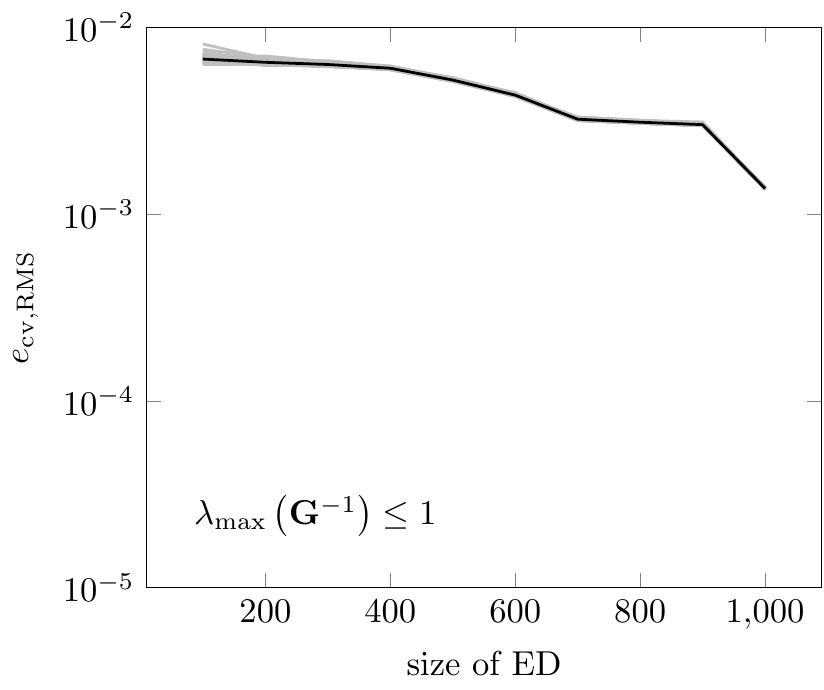}
		\hfill
		\includegraphics[width=0.32\textwidth]{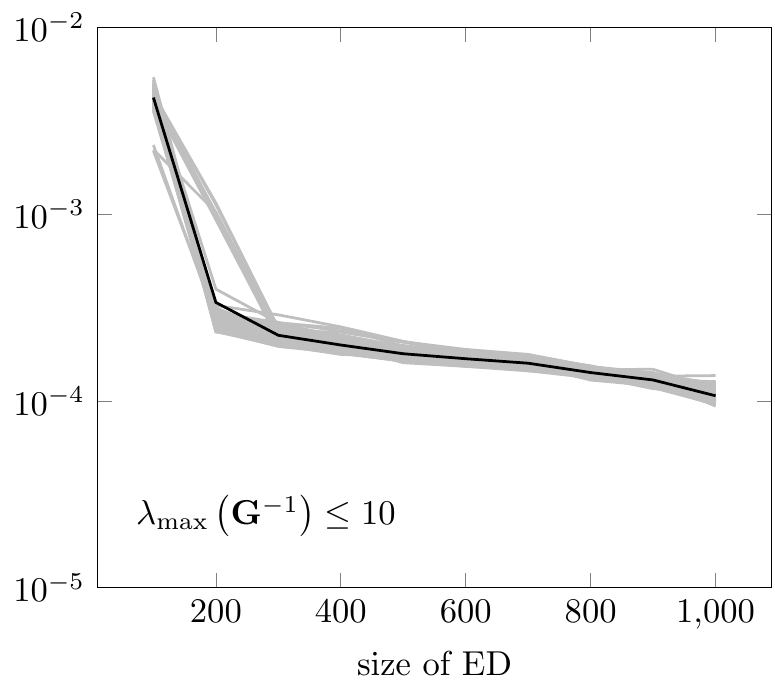}
		\hfill
		\includegraphics[width=0.32\textwidth]{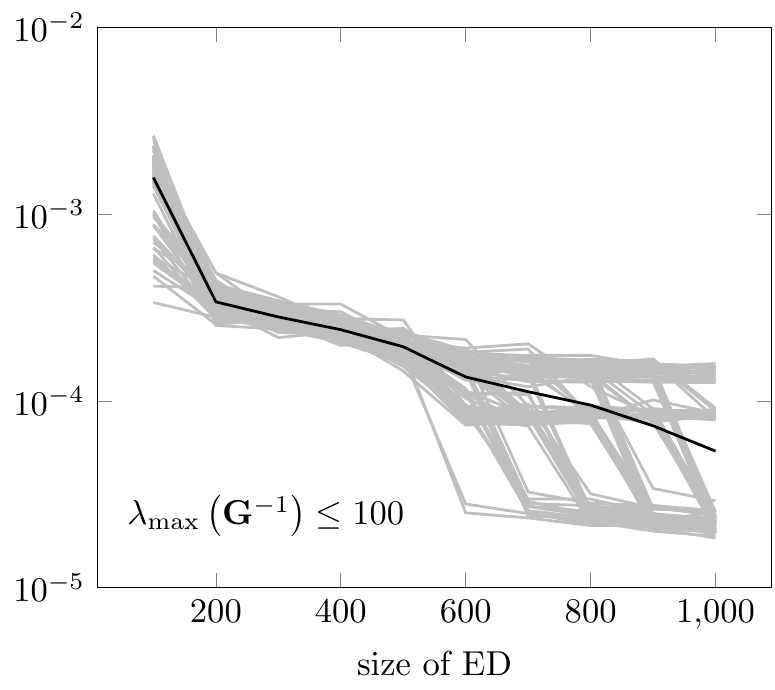}
	\end{center} 
	\caption{RMS CV errors of PCEs approximating the single-frequency dielectric-inset rectangular waveguide model, for EDs of increasing size and different sequential ED criteria. The gray lines show the results of 100 different EDs. The average errors are shown in black.}
	\label{fig:cv_rms_debye2_wg}
\end{figure}

In most \gls{uq} studies for frequency-response models, such as the waveguide examined here, one \gls{pce} per frequency point is developed in order to approximate the model's response over a frequency range.
Therefore, in this first numerical experiment, we will approximate the functional $r(\mathbf{y})$, given in decibels, for a fixed frequency $f=5$~GHz.
In particular, we employ Algorithm~\ref{algo:seq_ed} using 3 different condition number limits, 3 different maximum eigenvalue limits and 3 different maximum trace limits, with respect to the sequential expansion of the \gls{ed}. 
We construct \glspl{pce} using $100$ different random \glspl{ed}, i.e., using different random sampling seeds. 
The \gls{rms} \gls{cv} error \eqref{eq:cverr_rms} is computed using a random sample with $Q=10^5$ points.
The approximation results are shown in Figure~\ref{fig:cv_rms_debye2_wg}, where we omit the trace-related results, since they follow closely the eigenvalue-related ones.
Each subplot corresponds to a different sequential \gls{ed} criterion and shows the \gls{rms} \gls{cv} error of the \glspl{pce} for \glspl{ed} of increasing size.
In all cases, the surrogate models reach accuracies well beyond the ones typically needed in engineering applications.

Looking at the first two columns of Figure~\ref{fig:cv_rms_debye2_wg}, it can be indeed observed that more relaxed criteria improve the accuracy of the \gls{pce} on average.
Relaxing the condition-number criterion from $\kappa\left(\mathbf{D}\right) \leq \sqrt{3}$ to $\kappa\left(\mathbf{D}\right) \leq 10$ does not seem to introduce larger variations with respect to the \glspl{pce}' accuracy.
A more pronounced difference is observed between the eigenvalue-based criteria $\lambda_{\max} \leq 1$ and $\lambda_{\max} \leq 10$, however, the \glspl{pce} can in both cases be regarded as robust.
The rightmost columns of Figure~\ref{fig:cv_rms_debye2_wg} show that further relaxation of the criteria results in either marginal gains (bottom row), or in accuracy deterioration (top row).
Comparing the top and bottom-row sub-figures, the eigenvalue-based criteria seem to result in more robust results for a similar accuracy-cost relation.
In terms of a compromise between costs and accuracy, the best choices regarding the sequential expansion of the \gls{ed} are found to be $\kappa\left(\mathbf{D}\right) \leq 10$ and $\lambda_{\max} \leq 10$, for this particular model. 
However, we should note that \glspl{pce} based on the strictest and most robust criteria, i.e., $\kappa\left(\mathbf{D}\right) \leq \sqrt{3}$ and $\lambda_{\max} \leq 1$, also yield errors below the engineering standards, on top of being more robust in their results.

\subsubsection{Broadband Surrogate Modeling}
\label{subsubsec:debyw2_bb}
As mentioned in Section~\ref{subsubsec:debye2_sf}, the frequency response is typically approximated using one \gls{pce} per frequency point, equivalently, per time step in time-domain approaches \cite{xu2010, prasad2016}.
This can be computationally expensive in cases where a large number of frequency points must be examined.
For high-frequency models where the frequency dependence is also a relatively smooth functional, e.g., no sharp resonances exist in the examined frequency range, one can extend the presented surrogate modeling approach, such that it includes the frequency dependence as well \cite{georg2018}.
Specifically, while not random, the frequency can be modeled as a parameter which is uniformly distributed in the specified frequency range.
The resulting \gls{pce} approximates the functional $r(f,\mathbf{y})$, including the frequency dependence next to the geometrical and material parameters.

\begin{figure}[t]
	\begin{center}
		\includegraphics[width=0.339\textwidth]{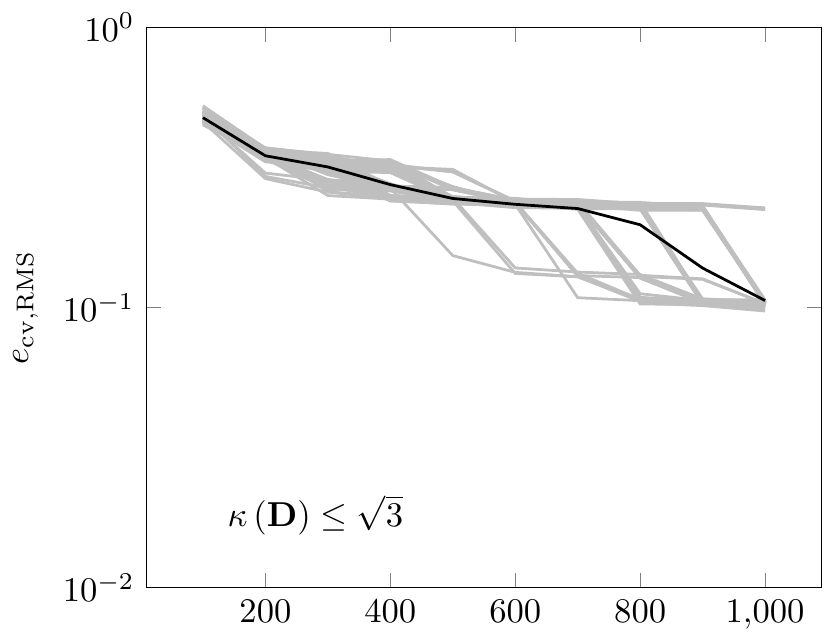} 
		\hfill
		\includegraphics[width=0.32\textwidth]{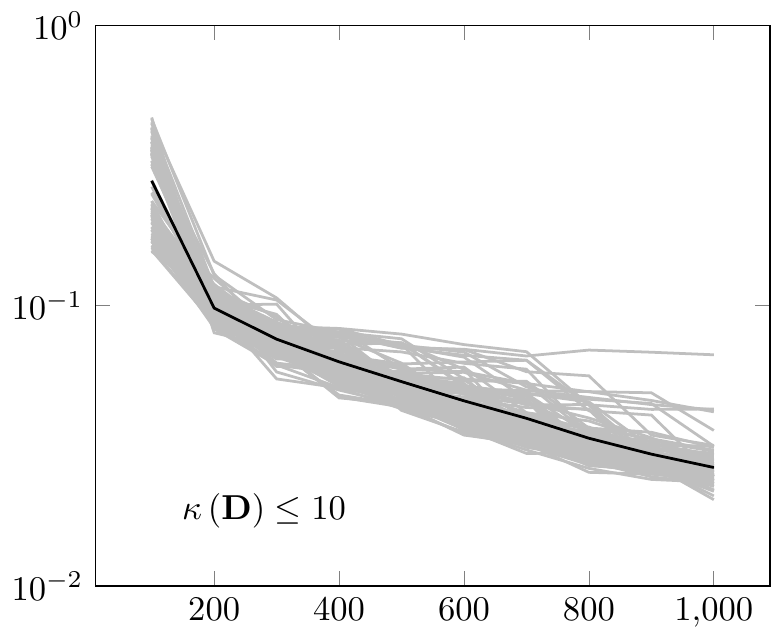}
		\hfill
		\includegraphics[width=0.32\textwidth]{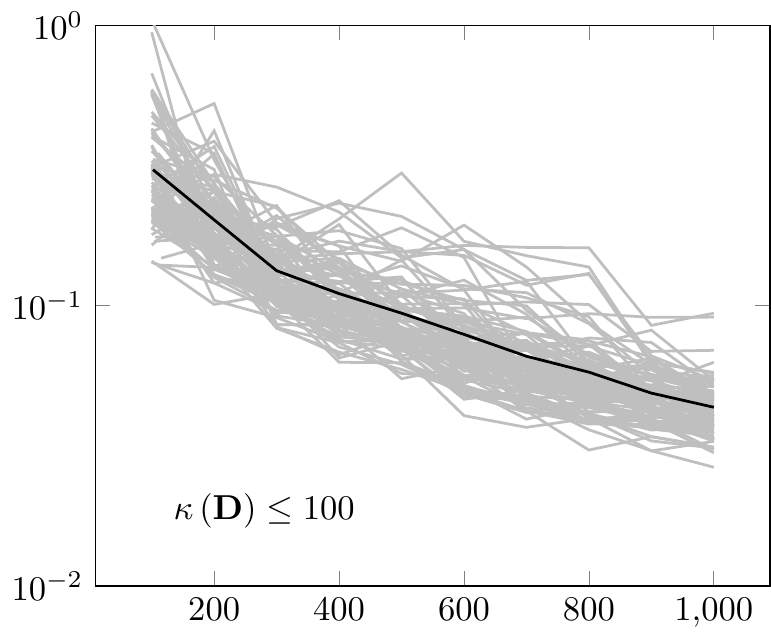} 
		\\
		\includegraphics[width=0.338\textwidth]{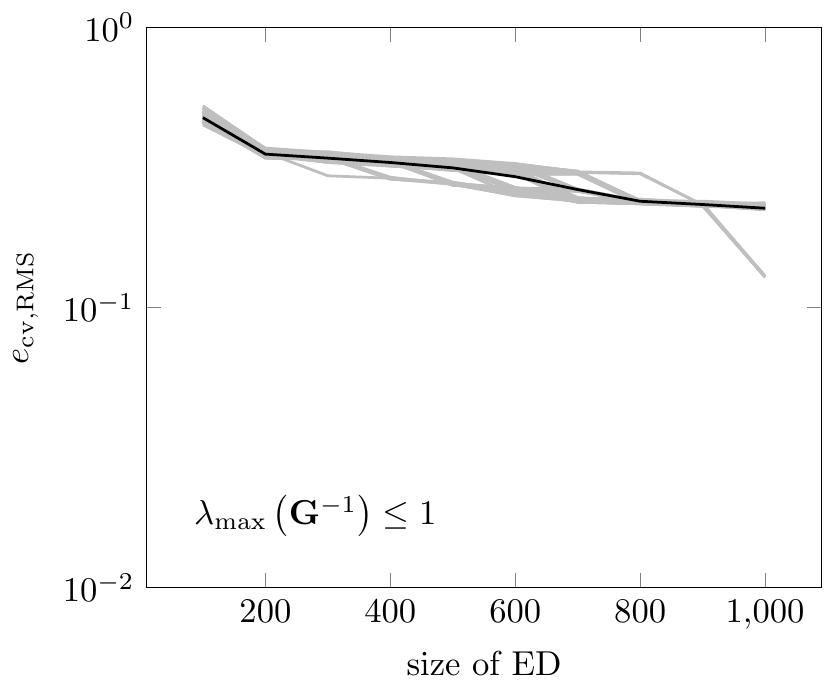}
		\hfill
		\includegraphics[width=0.32\textwidth]{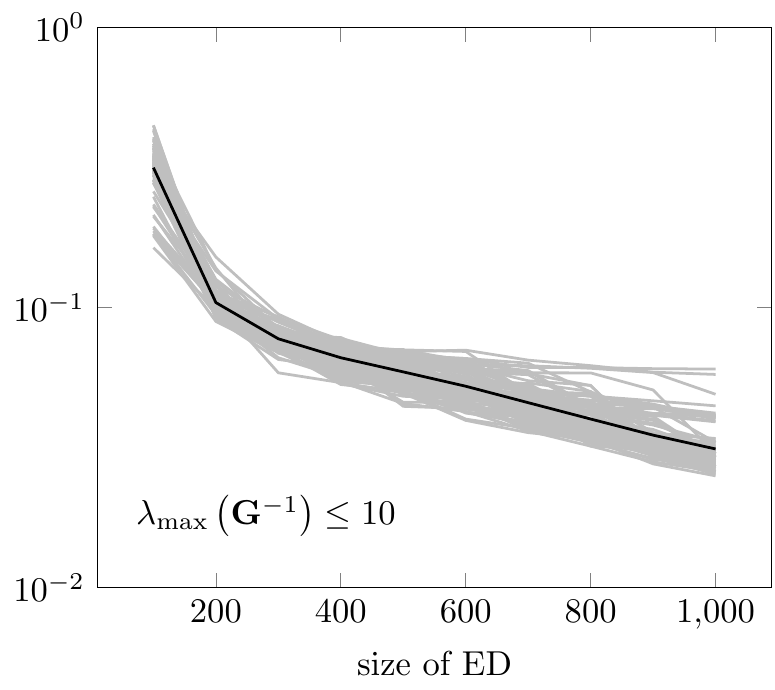}
		\hfill
		\includegraphics[width=0.32\textwidth]{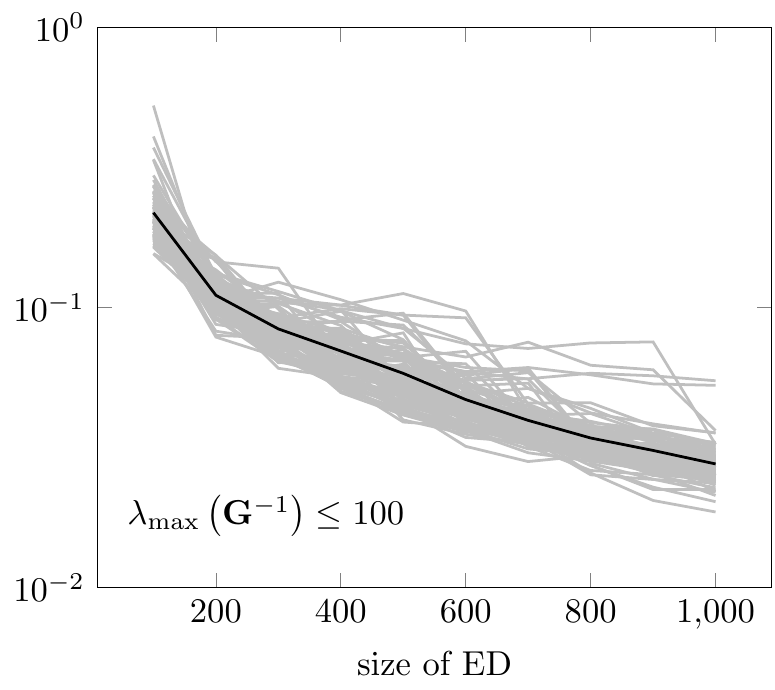}
	\end{center} 
	\caption{RMS CV errors of PCEs approximating the broadband dielectric-inset rectangular waveguide model, for EDs of increasing size and different sequential ED criteria. The gray lines show the results of 100 different EDs. The average errors are shown in black.}
	\label{fig:cv_rms_debye2_wg_bb}
\end{figure}

Using this idea, we repeat the numerical experiments of Section~\ref{subsubsec:debye2_sf}, where the frequency is now an additional uniformly distributed parameter in the range $\left[10, 20\right]$~GHz.
The results are shown in Figure~\ref{fig:cv_rms_debye2_wg_bb}.
The surrogate models reach satisfactory accuracies for the whole frequency range.
Similar to the single-frequency case, the eigenvalue-based criteria seem to be more robust compared to the condition-number-based ones, for a similar accuracy-cost relation.
Once more, the conditions $\kappa\left(\mathbf{D}\right) \leq 10$ and $\lambda_{\max}\left(\mathbf{G}^{-1}\right) \leq 10$ yield the best compromises between approximation accuracy and robustness.
In both cases, further relaxation of the sequential \gls{ed} criteria not only adds significant variation in the results, but also worsens the average accuracy.

\subsection{Optical Grating Coupler}
For the next, more challenging numerical example, we use a $1$D grating coupler model \cite{pitelet2019}. 
Such nanometer-scaled devices are employed in the field of nano-photonics and plasmonics \cite{schuller2010, akselrod2014, akselrod2015}. 
Their multi-layer structure consists of a high index dielectric on top of a metallic grating, which is placed on a substrate. 
A simplified, schematic model is depicted in Figure~\ref{fig:grating_structure}. 
We assume the structure to be infinitely extended in the lateral directions. 
Furthermore, only normal incident light beams from the top are considered. 
For certain incident wavelengths, matching the geometry dimensions, surface plasmons are excited in the metallic grating. 
Those wavelengths can be detected by observing the reflection coefficient $r$, which shows a dip in the frequency response each time the condition is satisfied, i.e., when a surface plasmon is excited, see Figure~\ref{fig:grating_broadband}. 
Note that, due to the resonances, a polynomial approximation of the grating coupler model including the frequency-responce becomes significantly more challenging.

The surface plasmon coupling is highly sensitive to the coupler's geometrical parameters, in particular, the grating pitch length $d_\mathrm{G}$, the groove $a_\mathrm{G}$, the thickness of the metallic and dielectric layers $h_\mathrm{m}$, respectively, $h_\mathrm{D}$, as well as the grating thickness $h_\mathrm{G}$ (see Figure~\ref{fig:grating_structure}). 
A full parameter study can be found in \cite{pitelet2019}. 
Following the same work, we model those four parameters as uniformly distributed and independent \glspl{rv}, such that $Y_n \sim \mathcal{U}[\bar{y}_n - \delta_{y_n}, \bar{y}_n + \delta_{y_n}]$, where $\bar{y}_n$ denotes the nominal value and $\delta_{y_n}$ the maximum allowed deviation from the nominal value. 
The parameters, their nominal values, and the allowed deviations are shown in Table \ref{tab:uncertain_parameter_grating}. 

\subsubsection{Broadband Surrogate Modeling}
In this example, we only consider the broadband approximation of the  grating coupler model, similar to Section~\ref{subsubsec:debyw2_bb}.
Thus, we model the wavelength as uniformly distributed in $\left[\SI{550}{\nano\meter}, \SI{800}{\nano\meter}\right]$. 
For a given realization of the parameter vector $\mathbf{y} = \mathbf{Y}\left(\theta\right)$ and a wavelength value $\lambda$, the reflection coefficient is a deterministic \gls{qoi} denoted by $r(\lambda,\mathbf{y})$ and given in decibels.
For the computation of the reflection coefficient, we employ a \gls{rcwa} code \cite{granet1996, lalanne1996}. 

\begin{figure}[t]
	\minipage[b][][b]{0.48\textwidth}
	\centering
	\includegraphics{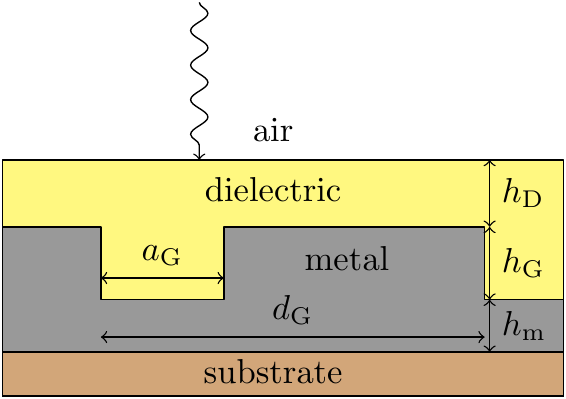} 
	\caption{Schematic view of the optical grating coupler.}
	\label{fig:grating_structure}
	\endminipage\hfill
	\minipage[b][][b]{0.48\textwidth}
	\includegraphics{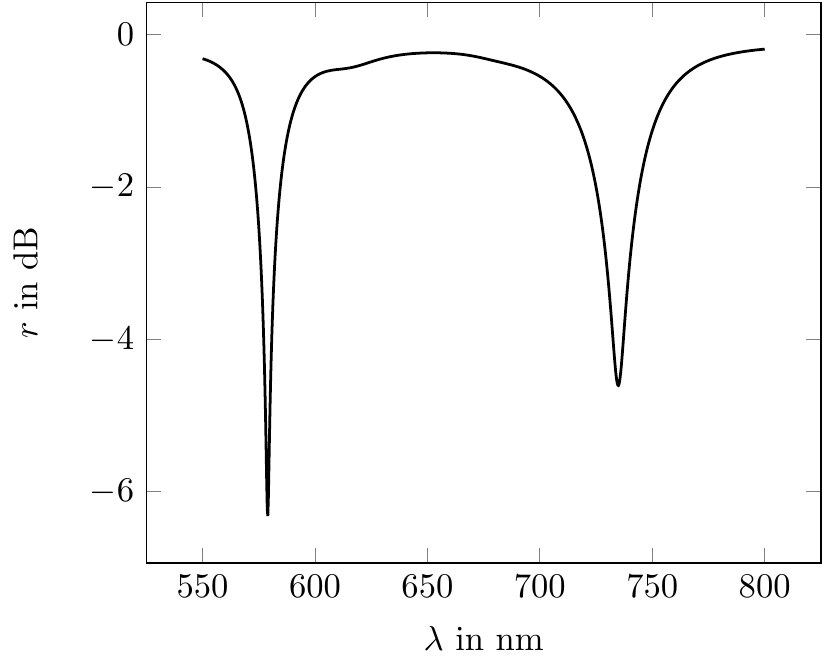} 
	\caption{Frequency-response of the optical grating coupler for the nominal geometry parameters.}
	\label{fig:grating_broadband}
	\endminipage\hfill		
\end{figure} 
\begin{table}[htb]
	\centering
	\caption{Uncertain geometric parameters of the  grating structure.} 
	\centering
	\raisebox{\depth}{\begin{tabular}[htbp]{@{}ccc@{}}
			\toprule
			parameter & mean $\bar{y} \left[\SI{}{\nano\meter}\right]$ & variation $\delta_y \left[\SI{}{\nano\meter}\right]$ \\
			\midrule
			$h_\mathrm{D}$   & $84.8$   & $0.3$\\
			$h_\mathrm{G}$   & $68.1$   & $0.1$\\ 
			$d_\mathrm{G}$   & $499.2$  & $1.0$ \\
			$a_\mathrm{G}$   & $165.4$  & $1.5$ \\  
			\bottomrule
	\end{tabular}}
	\label{tab:uncertain_parameter_grating}
\end{table}

\begin{figure}[t]
	\begin{center}
		\includegraphics[width=0.339\textwidth]{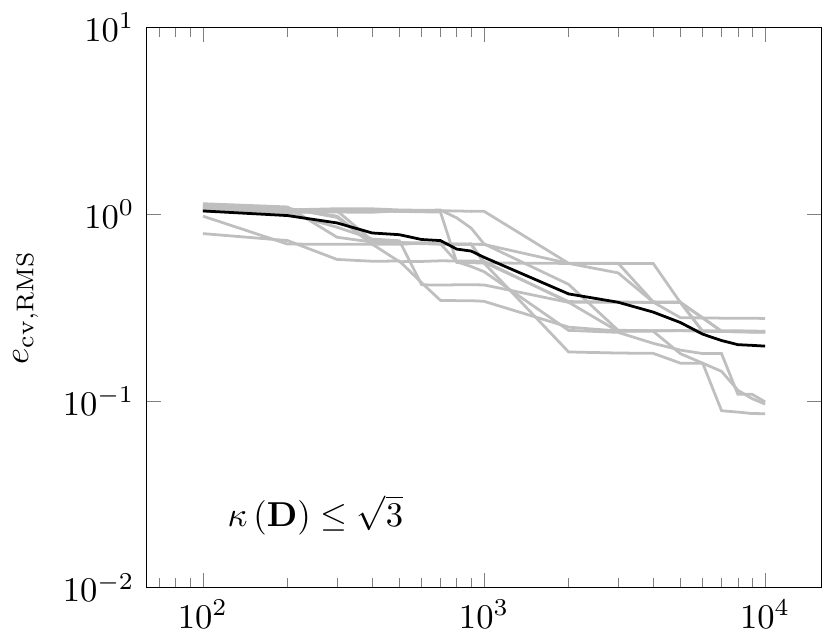}
		\hfill
		\includegraphics[width=0.32\textwidth]{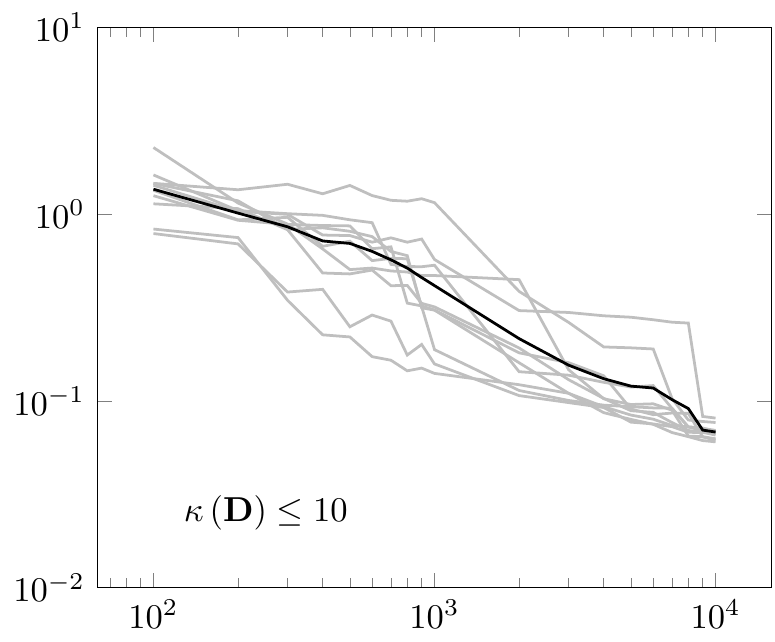}
		\hfill
		\includegraphics[width=0.32\textwidth]{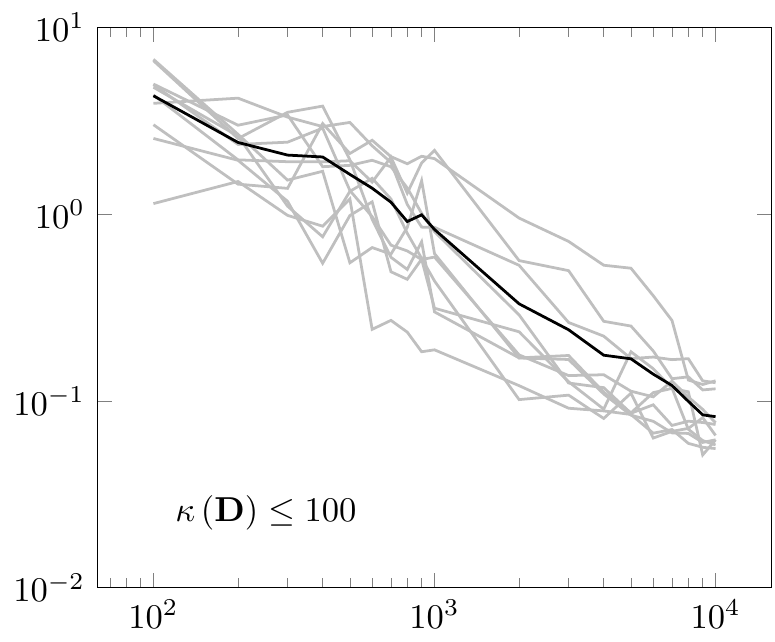} 
		\\
		\includegraphics[width=0.338\textwidth]{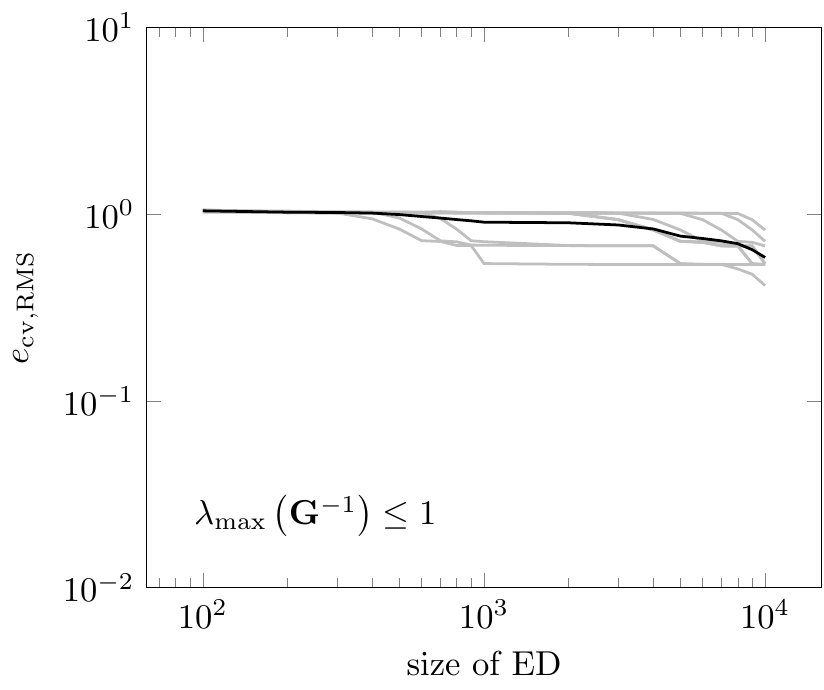}
		\hfill
		\includegraphics[width=0.32\textwidth]{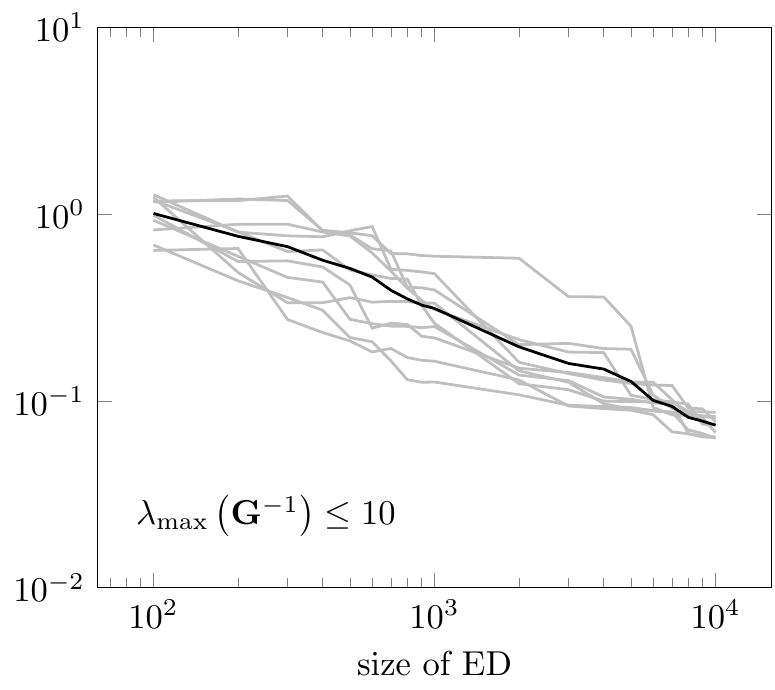}
		\hfill
		\includegraphics[width=0.32\textwidth]{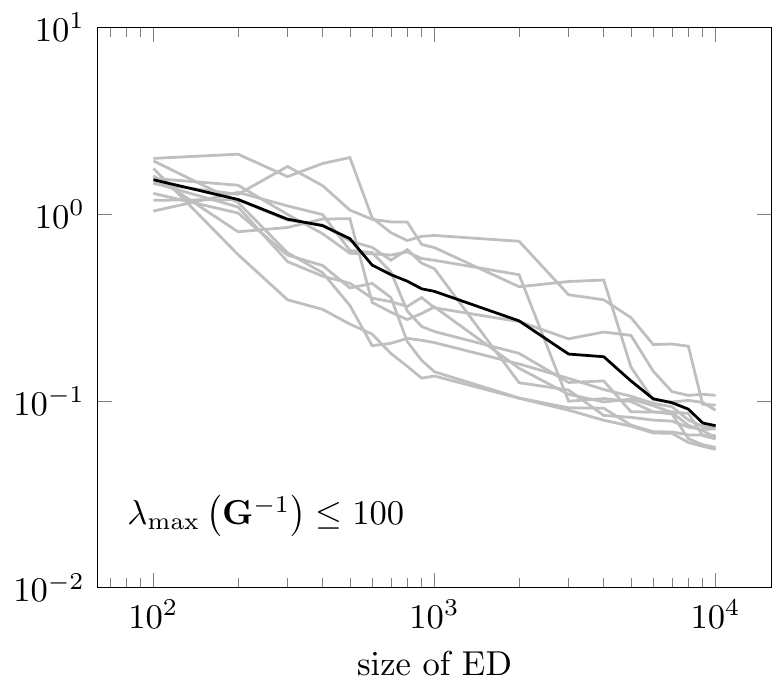}
	\end{center} 
	\caption{RMS CV errors of PCEs approximating the broadband grating coupler model, for EDs of increasing size and different sequential ED criteria. The gray lines show the results of 10 different EDs. The average errors are shown in black.}
	\label{fig:cv_rms_grating}
\end{figure}

Numerical results for different sequential \gls{ed} conditions are depicted in Figure~\ref{fig:cv_rms_grating}.
Once more, we omit the trace-based criterion, as the corresponding results are similar to the eigenvalue-based ones.
Despite the non-smooth frequency response, our method is able to provide accurate approximations, albeit at an elevated cost, i.e., \glspl{ed} with up to $10^4$ parameter realizations are employed.
It is worth stating that the resonances do not render \glspl{pce} altogether inapplicable, as is often the case for non-smooth responses.
Additionally considering the higher computational cost of the \gls{rcwa} solver, we employ only $10$ different \glspl{ed}. 
The size of the \gls{cv} sample used for computing the error \eqref{eq:cverr_rms} remains equal to $Q=10^5$.

Similar to the results of Section~\ref{subsec:diel_slab_wg}, relaxed sequential \gls{ed} criteria improve the average approximation accuracy, and at the same time introduce a larger variation among \glspl{pce} constructed for different \glspl{ed}, as can be observed from the first two columns of Figure~\ref{fig:grating_broadband}.
The last column of Figure~\ref{fig:grating_broadband} shows once more that, after a certain point, further relaxation of the sequential \gls{ed} criteria only enlarges the variation without accuracy improvement.
Once again, the eigenvalue-based criteria yield more robust results compared to the condition-number-based ones, however, the difference is not as pronounced as in Section~\ref{subsec:diel_slab_wg}.
Moreover, using the criterion $\lambda_{\max}(\mathbf{G}^{-1}) \leq 1$, almost no accuracy improvement is obtained for larger datasets.
Similar to Section~\ref{subsec:diel_slab_wg}, the conditions $\kappa(\mathbf{D}) \leq 10$ and $\lambda_{\max}(\mathbf{G}^{-1}) \leq 10$ provide the best trade-off between accuracy and robustness.  

\section{Summary and Conclusions}
\label{sec:conclusion}
In this work, we proposed an algorithm to construct sparse \gls{ls}-based \glspl{pce}. The algorithm features a sensitivity-based, adaptive selection of the polynomial basis terms, as well as a sequential \gls{ed} strategy, such that the available dataset of parameter realizations and \gls{qoi} observations is expanded at wish, given different optimality criteria. 
We focused on three such criteria, namely $K$-, $E$- and $A$-optimality, to construct nested datasets of \glspl{ed} and observations, and investigated the influence of different optimality conditions on the accuracy and robustness of the \gls{pce}-based surrogate models.

The method's accuracy and efficiency has been verified on two high-frequency electromagnetic models. 
Although comparisons between our approach and competitive methods, either non-adaptive or adaptive, have not been presented here, our method typically outperforms fixed-degree \gls{pce} approaches by orders and has been found to be superior to the popular \gls{lar}-\gls{pce} approach \cite{loukrezisPhD}.
Instead, this work focused on the largely unexplored topic of \gls{pce} robustness, which is closely related to the stability of the discrete \gls{ls} regression problem. 
In turn, \gls{ls} stability can be quantified by certain algebraic measures of the corresponding information matrix, with each of these measures being related to a so-called optimal \gls{ed} criterion.
It was repeatedly shown that strict criteria result in robust \glspl{pce}, the accuracy of which remains relatively unaffected by the given dataset.
Relaxing those criteria can improve the on-average accuracy at the cost of robustness, i.e., resulting in larger variations among \glspl{pce} constructed with different datasets. 
Relaxation of the sequential \gls{ed} criteria beyond a certain point only introduces more variation, while at the same time not improving or even worsening the average \gls{pce} accuracy.
In all numerical experiments, a good accuracy-robustness trade-off has been acquired with the criteria $\kappa\left(\mathbf{D}\right) \leq 10$ and $\lambda_{\max}\left(\mathbf{G}^{-1}\right) \leq 10$, the latter being typically more robust for approximations of similar accuracy. 
Compared to the theoretically optimal condition $\kappa\left(\mathbf{G}\right) \leq 3$ \cite{chkifa2015, cohen2013, migliorati2014, migliorati2015}, equivalently, $\kappa\left(\mathbf{D}\right) \leq \sqrt{3}$, our results show that those relaxed criteria yield significant accuracy gains, while keeping result variations at a modest level.

A continuation of the present work shall focus on developing surrogate models for high frequency \gls{em} applications featuring sharp resonances.
In such cases, most polynomial approximations will fail to accurately capture the frequency response.
A possible remedy could be found in multi-element \gls{pce} methods \cite{jakeman2013, wan2005, wan2006}, which are able to yield accurate approximations of non-smooth, discontinuous, or even singular parametric functions. 
This approach is currently under investigation and will be presented in a later study.

\subsection*{Acknowledgment}
D. Loukrezis and H. De Gersem would like to acknowledge the support of the Graduate School of Excellence for Computational Enginering at the Technische Universit\"at Darmstadt. D. Loukrezis further acknowledges the support of the BMBF via the research contract 05K19RDB. A. Galetzka's work is supported by the DFG through the Graduiertenkolleg 2128 ``Accelerator Science and Technology for Energy Recovery Linacs ''.

\bibliographystyle{abbrv}
\bibliography{references_alsace}



\end{document}

%% file: abbreviations.tex







\newacronym{ae}{a.e.}{almost every}
\newacronym{cad}{CAD}{computer-aided design}
\newacronym{cv}{CV}{cross-validation}
\newacronym{dc}{DC}{downward-closed}
\newacronym{dof}{DoF}{degrees of freedom}
\newacronym{em}{EM}{electromagnetic}
\newacronym{emf}{EMF}{electromagnetic field}
\newacronym{ed}{ED}{experimental design}
\newacronym{fdtd}{FDTD}{finite-difference time-domain}
\newacronym{fe}{FE}{finite element}
\newacronym{fem}{FEM}{finite element method}
\newacronym{gpc}{gPC}{generalized polynomial chaos}
\newacronym{hc}{HC}{hyperbolic cross}
\newacronym{hf}{HF}{high-frequency}
\newacronym{iga}{IGA}{isogeometric analysis}
\newacronym{iid}{iid}{independent and identically distributed}
\newacronym{loo}{LOO}{leave-one-out}
\newacronym{mim}{MIM}{metal-insulator-metal}
\newacronym{nurbs}{NURBS}{non-uniform rational B-splines}
\newacronym{omp}{OMP}{orthogonal matching pursuit}
\newacronym{pec}{PEC}{perfect electric conductor}
\newacronym{pmc}{PMC}{perfect magnetic conductor}
\newacronym{pml}{PML}{perfectly matched layers}
\newacronym{rf}{RF}{radio-frequency}
\newacronym{rhs}{RHS}{right-hand side}
\newacronym{rms}{RMS}{root-mean-square}
\newacronym{rcwa}{RCWA}{rigorous coupled wave analysis}
\newacronym{te}{TE}{transverse electric}
\newacronym{tm}{TM}{transverse magnetic}
\newacronym{tem}{TEM}{transverse electric and magnetic}
\newacronym{sqp}{SQP}{sequential quadratic programming}
\newacronym{svd}{SVD}{singular value decomposition}
\newacronym{lar}{LAR}{least angle regression}
\newacronym{lhs}{LHS}{latin hypercube sampling}
\newacronym{ls}{LS}{least squares}
\newacronym{mlmc}{MLMC}{multilevel Monte Carlo}
\newacronym{mc}{MC}{Monte Carlo}
\newacronym{pce}{PCE}{polynomial chaos expansion}
\newacronym{pde}{PDE}{partial differential equation}
\newacronym{pdf}{PDF}{probability density function}
\newacronym[plural=QoIs, firstplural=quantities of interest (QoIs)]{qoi}{QoI}{quantity of interest}
\newacronym{rv}{RV}{random variable}
\newacronym{rdp}{RDP}{random discrete projection}
\newacronym{qmc}{QMC}{quasi Monte Carlo}
\newacronym{td}{TD}{total degree}
\newacronym{tp}{TP}{tensor product}
\newacronym{uq}{UQ}{uncertainty quantification}



\newcommand{\AdmSet}{\Lambda^{\mathrm{adm}}}



\DeclareMathOperator*{\argmin}{arg\,min}
\DeclareMathOperator*{\argmax}{arg\,max}

\newcommand{\ImageSet}{\Xi}

%% file: inkscape/waveguide.pdf_tex
\begingroup%
  \makeatletter%
  \providecommand\color[2][]{%
    \errmessage{(Inkscape) Color is used for the text in Inkscape, but the package 'color.sty' is not loaded}%
    \renewcommand\color[2][]{}%
  }%
  \providecommand\transparent[1]{%
    \errmessage{(Inkscape) Transparency is used (non-zero) for the text in Inkscape, but the package 'transparent.sty' is not loaded}%
    \renewcommand\transparent[1]{}%
  }%
  \providecommand\rotatebox[2]{#2}%
  \newcommand*\fsize{\dimexpr\f@size pt\relax}%
  \newcommand*\lineheight[1]{\fontsize{\fsize}{#1\fsize}\selectfont}%
  \ifx\svgwidth\undefined%
    \setlength{\unitlength}{496.23214459bp}%
    \ifx\svgscale\undefined%
      \relax%
    \else%
      \setlength{\unitlength}{\unitlength * \real{\svgscale}}%
    \fi%
  \else%
    \setlength{\unitlength}{\svgwidth}%
  \fi%
  \global\let\svgwidth\undefined%
  \global\let\svgscale\undefined%
  \makeatother%
  \begin{picture}(1,1.07776395)%
    \lineheight{1}%
    \setlength\tabcolsep{0pt}%
    \put(0,0){\includegraphics[width=\unitlength,page=1]{inkscape/waveguide.pdf}}%
    \put(0.41520429,0.4858972){\color[rgb]{0,0,0}\makebox(0,0)[lt]{\lineheight{1.25}\smash{\begin{tabular}[t]{l}$\Gamma_\text{in}$\end{tabular}}}}%
    \put(0,0){\includegraphics[width=\unitlength,page=2]{inkscape/waveguide.pdf}}%
    \put(0.52910636,0.82246668){\color[rgb]{0,0,0}\makebox(0,0)[lt]{\lineheight{1.25}\smash{\begin{tabular}[t]{l}$\varepsilon,\,\mu$\end{tabular}}}}%
    \put(0.34453683,0.01065648){\color[rgb]{0,0,0}\makebox(0,0)[lt]{\lineheight{1.25}\smash{\begin{tabular}[t]{l}$w$\end{tabular}}}}%
    \put(-0.00055619,0.25512674){\color[rgb]{0,0,0}\makebox(0,0)[lt]{\lineheight{1.25}\smash{\begin{tabular}[t]{l}$h$\end{tabular}}}}%
    \put(0.57005781,0.13233335){\color[rgb]{0,0,0}\makebox(0,0)[lt]{\lineheight{1.25}\smash{\begin{tabular}[t]{l}$x$\end{tabular}}}}%
    \put(0.71844476,0.24419456){\color[rgb]{0,0,0}\makebox(0,0)[lt]{\lineheight{1.25}\smash{\begin{tabular}[t]{l}$z$\end{tabular}}}}%
    \put(0.62458382,0.24076383){\color[rgb]{0,0,0}\makebox(0,0)[lt]{\lineheight{1.25}\smash{\begin{tabular}[t]{l}$y$\end{tabular}}}}%
    \put(0.82645101,0.21494626){\color[rgb]{0,0,0}\makebox(0,0)[lt]{\lineheight{1.25}\smash{\begin{tabular}[t]{l}$d$\end{tabular}}}}%
    \put(0.9408425,0.4617694){\color[rgb]{0,0,0}\makebox(0,0)[lt]{\lineheight{1.25}\smash{\begin{tabular}[t]{l}$\ell$\end{tabular}}}}%
    \put(0,0){\includegraphics[width=\unitlength,page=3]{inkscape/waveguide.pdf}}%
  \end{picture}%
\endgroup%

%% file: alsace_paper_arXiv.bbl
\begin{thebibliography}{10}

\bibitem{akselrod2014}
G.~M. Akselrod, C.~Argyropoulos, T.~B. Hoang, C.~Cirac{\`i}, C.~Fang, J.~Huang,
  D.~R. Smith, and M.~H. Mikkelsen.
\newblock Probing the mechanisms of large purcell enhancement in plasmonic
  nanoantennas.
\newblock {\em Nature Photonics}, 8:835 EP --, Oct 2014.

\bibitem{akselrod2015}
G.~M. Akselrod, T.~Ming, C.~Argyropoulos, T.~B. Hoang, Y.~Lin, X.~Ling, D.~R.
  Smith, J.~Kong, and M.~H. Mikkelsen.
\newblock Leveraging nanocavity harmonics for control of optical processes in
  2d semiconductors.
\newblock {\em Nano Letters}, 15(5):3578--3584, 2015.
\newblock PMID: 25914964.

\bibitem{arras2019}
B.~Arras, M.~Bachmayr, and A.~Cohen.
\newblock Sequential sampling for optimal weighted least squares approximations
  in hierarchical spaces.
\newblock {\em SIAM Journal on Mathematics of Data Science}, 1(1):189--207,
  2019.

\bibitem{babuska2010}
I.~Babu\v{s}ka, F.~Nobile, and R.~Tempone.
\newblock A stochastic collocation method for elliptic partial differential
  equations with random input data.
\newblock {\em SIAM Review}, 52(2):317--355, 2010.

\bibitem{baudin2017}
M.~Baudin, A.~Dutfoy, B.~Iooss, and A.-L. Popelin.
\newblock {OpenTURNS}: An industrial software for uncertainty quantification in
  simulation.
\newblock In R.~Ghanem, D.~Higdon, and H.~Owhadi, editors, {\em Handbook of
  Uncertainty Quantification}, pages 2001--2038. Springer International
  Publishing, 2017.

\bibitem{bellman1957}
R.~E. Bellman.
\newblock {\em {Dynamic Programming}}.
\newblock Princeton University Press, 1957.

\bibitem{blatman2010}
G.~Blatman and B.~Sudret.
\newblock {An adaptive algorithm to build up sparse polynomial chaos expansions
  for stochastic finite element analysis}.
\newblock {\em Probabilistic Engineering Mechanics}, 25(2):183--197, 2010.

\bibitem{blatman2010a}
G.~Blatman and B.~Sudret.
\newblock Efficient computation of global sensitivity indices using sparse
  polynomial chaos expansions.
\newblock {\em Reliability Engineering and System Safety}, 95(11):1216 -- 1229,
  2010.

\bibitem{blatman2011}
G.~Blatman and B.~Sudret.
\newblock {Adaptive Sparse Polynomial Chaos Expansion Based on Least Angle
  Regression}.
\newblock {\em Journal of Computational Physics}, 230(6):2345--2367, 2011.

\bibitem{burnaev2017}
E.~Burnaev, I.~Panin, and B.~Sudret.
\newblock Efficient design of experiments for sensitivity analysis based on
  polynomial chaos expansions.
\newblock {\em Annals of Mathematics and Artificial Intelligence},
  81(1):187--207, Oct 2017.

\bibitem{caflisch1998}
R.~E. Caflisch.
\newblock {Monte Carlo} and quasi-{Monte Carlo} methods.
\newblock {\em Acta Numerica}, 7:1 -- 49, 1998.

\bibitem{chkifa2015}
A.~Chkifa, A.~Cohen, G.~Migliorati, F.~Nobile, and R.~Tempone.
\newblock Discrete least squares polynomial approximation with random
  evaluations - application to parametric and stochastic elliptic pdes.
\newblock {\em ESAIM: M2AN}, 49(3):815--837, 2015.

\bibitem{cohen2013}
A.~Cohen, M.~A. Davenport, and D.~Leviatan.
\newblock On the stability and accuracy of least squares approximations.
\newblock {\em Foundations of Computational Mathematics}, 13(5):819--834, 2013.

\bibitem{cohen2017}
A.~Cohen and G.~Migliorati.
\newblock Optimal weighted least-squares methods.
\newblock {\em SMAI Journal of Computational Mathematics}, 3:181--203, 2017.

\bibitem{cohen2018}
A.~Cohen and G.~Migliorati.
\newblock Multivariate approximation in downward closed polynomial spaces.
\newblock In J.~Dick, F.~Y. Kuo, and H.~Wo{\'{z}}niakowski, editors, {\em
  Contemporary Computational Mathematics - A Celebration of the 80th Birthday
  of Ian Sloan, Springer International Publishing}, pages 233--282, 2018.

\bibitem{diaz2018}
P.~Diaz, A.~Doostan, and J.~Hampton.
\newblock Sparse polynomial chaos expansions via compressed sensing and
  d-optimal design.
\newblock {\em Computer Methods in Applied Mechanics and Engineering}, 336:640
  -- 666, 2018.

\bibitem{doostan2011}
A.~Doostan and H.~Owhadi.
\newblock A non-adapted sparse approximation of pdes with stochastic inputs.
\newblock {\em Journal of Computational Physics}, 230(8):3015 -- 3034, 2011.

\bibitem{doostan2013}
A.~Doostan, A.~Validi, and G.~Iaccarino.
\newblock Non-intrusive low-rank separated approximation of high-dimensional
  stochastic models.
\newblock {\em Computer Methods in Applied Mechanics and Engineering}, 263:42
  -- 55, 2013.

\bibitem{fajraoui2017}
N.~Fajraoui, S.~Marelli, and B.~Sudret.
\newblock Sequential design of experiment for sparse polynomial chaos
  expansions.
\newblock {\em SIAM/ASA Journal on Uncertainty Quantification},
  5(1):1061--1085, 2017.

\bibitem{feinberg2018}
J.~Feinberg, V.~Eck, and H.~Langtangen.
\newblock Multivariate polynomial chaos expansions with dependent variables.
\newblock {\em SIAM Journal on Scientific Computing}, 40(1):A199--A223, 2018.

\bibitem{georg2018}
N.~Georg, D.~Loukrezis, U.~R{\"{o}}mer, and S.~Sch{\"{o}}ps.
\newblock Uncertainty quantification for an optical grating coupler with an
  adjoint-based leja adaptive collocation method.
\newblock {\em CoRR}, abs/1807.07485, 2018.

\bibitem{gerstner2003}
T.~Gerstner and M.~Griebel.
\newblock {Dimension-Adaptive Tensor-Product Quadrature}.
\newblock {\em Computing}, 71(1):65--87, 2003.

\bibitem{ghanem1991}
R.~G. Ghanem and P.~D. Spanos.
\newblock {\em {Stochastic Finite Elements: A Spectral Approach}}.
\newblock Springer-Verlag New York, Inc, New York, NY, USA, 1991.

\bibitem{gladwin2019}
K.~T. {Gladwin Jos} and K.~J. {Vinoy}.
\newblock A fast polynomial chaos expansion for uncertainty quantification in
  stochastic electromagnetic problems.
\newblock {\em IEEE Antennas and Wireless Propagation Letters}, pages 1--1,
  2019.

\bibitem{golub1996}
G.~H. Golub and C.~F. Van~Loan.
\newblock {\em Matrix Computations (3rd Ed.)}.
\newblock Johns Hopkins University Press, Baltimore, MD, USA, 1996.

\bibitem{granet1996}
G.~Granet and B.~Guizal.
\newblock Efficient implementation of the coupled-wave method for metallic
  lamellar gratings in tm polarization.
\newblock {\em Journal of the Optical Society of America A}, 13(5):1019--1023,
  May 1996.

\bibitem{hadigol2018}
M.~Hadigol and A.~Doostan.
\newblock Least squares polynomial chaos expansion: A review of sampling
  strategies.
\newblock {\em Computer Methods in Applied Mechanics and Engineering}, 332:382
  -- 407, 2018.

\bibitem{hampton2015}
J.~Hampton and A.~Doostan.
\newblock Coherence motivated sampling and convergence analysis of least
  squares polynomial chaos regression.
\newblock {\em Computer Methods in Applied Mechanics and Engineering}, 290:73
  -- 97, 2015.

\bibitem{hampton2015a}
J.~Hampton and A.~Doostan.
\newblock Compressive sampling of polynomial chaos expansions: Convergence
  analysis and sampling strategies.
\newblock {\em Journal of Computational Physics}, 280:363 -- 386, 2015.

\bibitem{higham2002}
N.~J. Higham.
\newblock {\em Accuracy and Stability of Numerical Algorithms}.
\newblock Society for Industrial and Applied Mathematics, second edition, 2002.

\bibitem{hu2018}
R.~{Hu}, V.~{Monebhurrun}, R.~{Himeno}, H.~{Yokota}, and F.~{Costen}.
\newblock An adaptive least angle regression method for uncertainty
  quantification in fdtd computation.
\newblock {\em IEEE Transactions on Antennas and Propagation},
  66(12):7188--7197, Dec 2018.

\bibitem{jakeman2015a}
J.~D. Jakeman, M.~S. Eldred, and K.~Sargsyan.
\newblock Enhancing $\ell_1$-minimization estimates of polynomial chaos
  expansions using basis selection.
\newblock {\em Journal of Computational Physics}, 289:18 -- 34, 2015.

\bibitem{jakeman2013}
J.~D. Jakeman, A.~Narayan, and D.~Xiu.
\newblock Minimal multi-element stochastic collocation for uncertainty
  quantification of discontinuous functions.
\newblock {\em Journal of Computational Physics}, 242:790 -- 808, 2013.

\bibitem{jankoski2019}
R.~Jankoski, U.~R{\"o}mer, and S.~Sch{\"o}ps.
\newblock Stochastic modeling of magnetic hysteretic properties by using
  multivariate random fields.
\newblock {\em International Journal for Uncertainty Quantification},
  9(1):85--102, 2019.

\bibitem{konakli2016a}
K.~Konakli and B.~Sudret.
\newblock Global sensitivity analysis using low-rank tensor approximations.
\newblock {\em Reliability Engineering \& System Safety}, 156:64 -- 83, 2016.

\bibitem{konakli2016}
K.~Konakli and B.~Sudret.
\newblock Polynomial meta-models with canonical low-rank approximations:
  Numerical insights and comparison to sparse polynomial chaos expansions.
\newblock {\em Journal of Computational Physics}, 321:1144 -- 1169, 2016.

\bibitem{lalanne1996}
P.~Lalanne and G.~M. Morris.
\newblock Highly improved convergence of the coupled-wave method for tm
  polarization.
\newblock {\em Journal of the Optical Society of America A}, 13(4):779--784,
  Apr 1996.

\bibitem{lebrun2009}
R.~Lebrun and A.~Dutfoy.
\newblock A generalization of the nataf transformation to distributions with
  elliptical copula.
\newblock {\em Probabilistic Engineering Mechanics}, 24(2):172 -- 178, 2009.

\bibitem{lemieux2009}
C.~Lemieux.
\newblock {\em {Monte Carlo} and Quasi-{Monte Carlo} Sampling}.
\newblock {Springer Series in Statistics}. 2009.

\bibitem{loukrezisPhD}
D.~Loukrezis.
\newblock {\em Adaptive approximations for high-dimensional uncertainty
  quantification in stochastic parametric electromagnetic field simulations}.
\newblock PhD thesis, Technische Universit{\"a}t Darmstadt, 2019.

\bibitem{loukrezis2019}
D.~Loukrezis, U.~R{\"o}mer, and H.~De~Gersem.
\newblock Assessing the performance of {Leja} and {Clenshaw-Curtis} collocation
  for computational electromagnetics with random input data.
\newblock {\em International Journal for Uncertainty Quantification},
  9(1):33--57, 2019.

\bibitem{migliorati2013a}
G.~Migliorati.
\newblock {Adaptive Polynomial Approximation by Means of Random Discrete Least
  Squares}.
\newblock In A.~Abdulle, S.~Deparis, D.~Kressner, F.~Nobile, and M.~Picasso,
  editors, {\em {ENUMATH}}, volume 103 of {\em {Lecture Notes in Computational
  Science and Engineering}}, pages 547--554. Springer, 2013.

\bibitem{migliorati2015a}
G.~Migliorati.
\newblock Learning with discrete least squares on multivariate polynomial
  spaces using evaluations at random or low-discrepancy point sets.
\newblock In P.~Pardalos, M.~Pavone, G.~M. Farinella, and V.~Cutello, editors,
  {\em Machine Learning, Optimization, and Big Data}, pages 1--13. Springer
  International Publishing, 2015.

\bibitem{migliorati2015}
G.~Migliorati and F.~Nobile.
\newblock Analysis of discrete least squares on multivariate polynomial spaces
  with evaluations at low-discrepancy point sets.
\newblock {\em Journal of Complexity}, 31(4):517--542, 2015.

\bibitem{migliorati2013}
G.~Migliorati, F.~Nobile, E.~{von~Schwerin}, and R.~Tempone.
\newblock {Approximation of Quantities of Interest in Stochastic PDEs by the
  Random Discrete L2 Projection on Polynomial Spaces}.
\newblock {\em SIAM Journal on Scientific Computing}, 35(3), 2013.

\bibitem{migliorati2014}
G.~Migliorati, F.~Nobile, E.~{von~Schwerin}, and R.~Tempone.
\newblock {Analysis of Discrete L2 Projection on Polynomial Spaces with Random
  Evaluations}.
\newblock {\em Foundations of Computational Mathematics}, 14(3):419--456, 2014.

\bibitem{nguyen2016}
T.~T. {Nguyen}, D.~H. {Mac}, and S.~{Cl{\'e}net}.
\newblock Uncertainty quantification using sparse approximation for models with
  a high number of parameters: Application to a magnetoelectric sensor.
\newblock {\em IEEE Transactions on Magnetics}, 52(3):1--4, March 2016.

\bibitem{offermann2015}
P.~{Offermann}, H.~{Mac}, T.~T. {Nguyen}, S.~{Cl{\'e}net}, H.~{De Gersem}, and
  K.~{Hameyer}.
\newblock Uncertainty quantification and sensitivity analysis in electrical
  machines with stochastically varying machine parameters.
\newblock {\em IEEE Transactions on Magnetics}, 51(3):1--4, March 2015.

\bibitem{oladyshkin2012}
S.~Oladyshkin and W.~Nowak.
\newblock Data-driven uncertainty quantification using the arbitrary polynomial
  chaos expansion.
\newblock {\em Reliability Engineering \& System Safety}, 106:179--190, 2012.

\bibitem{peng2014}
J.~Peng, J.~Hampton, and A.~Doostan.
\newblock A weighted l1-minimization approach for sparse polynomial chaos
  expansions.
\newblock {\em Journal of Computational Physics}, 267:92 -- 111, 2014.

\bibitem{pitelet2019}
A.~Pitelet, N.~Schmitt, D.~Loukrezis, C.~Scheid, H.~D. Gersem, C.~Cirac\`{i},
  E.~Centeno, and A.~Moreau.
\newblock Influence of spatial dispersion on surface plasmons, nanoparticles,
  and grating couplers.
\newblock {\em Journal of the Optical Society of America B}, 36(11):2989--2999,
  Nov 2019.

\bibitem{prasad2016}
A.~K. {Prasad}, M.~{Ahadi}, and S.~{Roy}.
\newblock Multidimensional uncertainty quantification of microwave/rf networks
  using linear regression and optimal design of experiments.
\newblock {\em IEEE Transactions on Microwave Theory and Techniques},
  64(8):2433--2446, Aug 2016.

\bibitem{schuller2010}
J.~A. Schuller, E.~S. Barnard, W.~Cai, Y.~C. Jun, J.~S. White, and M.~L.
  Brongersma.
\newblock Plasmonics for extreme light concentration and manipulation.
\newblock {\em Nature Materials}, 9(3):193--204, 2010.

\bibitem{smith2014}
R.~Smith.
\newblock {\em {Uncertainty Quantification: Theory, Implementation, and
  Applications}}.
\newblock Computational Science and Engineering, SIAM. 2014.

\bibitem{soize2004}
C.~Soize and R.~Ghanem.
\newblock Physical systems with random uncertainties: Chaos representations
  with arbitrary probability measure.
\newblock {\em SIAM Journal on Scientific Computing}, 26(2):395--410, 2004.

\bibitem{sudret2008}
B.~Sudret.
\newblock Global sensitivity analysis using polynomial chaos expansions.
\newblock {\em Reliability Engineering \& System Safety}, 93(7):964 -- 979,
  2008.

\bibitem{sullivan2015}
T.~J. Sullivan.
\newblock {\em {Introduction to Uncertainty Quantification}}.
\newblock Texts in Applied Mathematics, Springer International Publishing
  Switzerland. 2015.

\bibitem{wan2006}
X.~Wan and G.~Karniadakis.
\newblock {Multi-Element Generalized Polynomial Chaos for Arbitrary Probability
  Measures}.
\newblock {\em SIAM Journal on Scientific Computing}, 28(3):901--928, 2006.

\bibitem{wan2005}
X.~Wan and G.~E. Karniadakis.
\newblock An adaptive multi-element generalized polynomial chaos method for
  stochastic differential equations.
\newblock {\em Journal of Computational Physics}, 209(2):617 -- 642, 2005.

\bibitem{wan2006a}
X.~Wan and G.~E. Karniadakis.
\newblock Beyond {Wiener}--{Askey} expansions: Handling arbitrary {PDFs}.
\newblock {\em Journal of Scientific Computing}, 27(1):455--464, Jun 2006.

\bibitem{xiu2002}
D.~Xiu and G.~E. Karniadakis.
\newblock {{The Wiener-Askey Polynomial Chaos for Stochastic Differential
  Equations}}.
\newblock {\em SIAM Journal on Scientific Computing}, 24(2):619--644, 2002.

\bibitem{xu2010}
J.~{Xu}, M.~Y. {Koledintseva}, Y.~{Zhang}, Y.~{He}, B.~{Matlin}, R.~E.
  {DuBroff}, J.~L. {Drewniak}, and J.~{Zhang}.
\newblock Complex permittivity and permeability measurements and
  finite-difference time-domain simulation of ferrite materials.
\newblock {\em IEEE Transactions on Electromagnetic Compatibility},
  52(4):878--887, Nov 2010.

\end{thebibliography}
